\documentclass[sigconf,screen]{acmart}

\AtBeginDocument{%
  }

\copyrightyear{2024}
\acmYear{2024}
\setcopyright{acmlicensed}\acmConference[ESEM '24]{Proceedings of the 18th ACM / IEEE International Symposium on Empirical Software Engineering and Measurement}{October 24--25, 2024}{Barcelona, Spain}
\acmBooktitle{Proceedings of the 18th ACM / IEEE International Symposium on Empirical Software Engineering and Measurement (ESEM '24), October 24--25, 2024, Barcelona, Spain}
\acmDOI{10.1145/3674805.3686685}
\acmISBN{979-8-4007-1047-6/24/10}

\usepackage{subcaption}
\usepackage{listings}
\usepackage{pythonhighlight}
\usepackage{courier}

\definecolor{codegreen}{rgb}{0,0.6,0}
\definecolor{codegray}{rgb}{0.5,0.5,0.5}
\definecolor{codepurple}{rgb}{0.58,0,0.82}
\definecolor{backcolour}{rgb}{1,1,1}
\definecolor{deleted}{rgb}{0.8, 0.0, 0.0}
\definecolor{added}{rgb}{0.0, 0.9, 0.2}

\lstdefinestyle{pythoncode}{
	language=Python,
	backgroundcolor=\color{backcolour},   
	commentstyle=\color{codegreen},
	keywordstyle=\color{orange},
	numberstyle=\tiny\color{codegray},
	stringstyle=\color{codepurple},
	basicstyle=\fontfamily{lmss}\tiny,
	breakatwhitespace=false,         
	breaklines=true,                 
	captionpos=b,                    
	keepspaces=true,                 
	numbers=left,                    
	numbersep=5pt,
	xleftmargin=0.3cm,                  
	showspaces=false,                
	showstringspaces=false,
	showtabs=false,                  
	tabsize=2,
	frame=top,
	frame=bottom,
	morecomment=[f][\color{deleted}]{-},
    morecomment=[f][\color{added}]{+}
}
\usepackage{multirow}
\usepackage{booktabs}
\usepackage{tablefootnote}
\usepackage{caption}
\usepackage{enumitem}
\usepackage{colortbl}
\usepackage{xspace}

\usepackage{url}

\usepackage[absolute,overlay]{textpos}
\usepackage{tikz}
\usepackage{xcolor}
\usepackage{calc}

\newcounter{findingctr}

\newcommand{\finding}[2]{\refstepcounter{findingctr}\emph{#1:\label{box:#2}}}

\newlength{\boxw}
\newlength{\boxh}
\newlength{\shadowsize}
\newlength{\boxroundness}
\newlength{\tmpa}
\newsavebox{\shadowblockbox}

\newenvironment{findingenv}[2]%
{\vspace{0.2cm}\noindent
\begin{lrbox}{
\shadowblockbox
}
\begin{minipage}{.98\columnwidth}
\finding{#1}{#2}~}%
{\end{minipage}\end{lrbox}%
\settowidth{\boxw}{\usebox{\shadowblockbox}}%
\settodepth{\tmpa}{\usebox{\shadowblockbox}}%
\settoheight{\boxh}{\usebox{\shadowblockbox}}%
\addtolength{\boxh}{\tmpa}%
\begin{tikzpicture}
\addtolength{\boxw}{\boxroundness * 2}
\addtolength{\boxh}{\boxroundness * 2}

\foreach \x in {0,.05,...,1}
{
\setlength{\tmpa}{\shadowsize * \real{\x}}
\fill[xshift=\shadowsize - 1pt,yshift=-\shadowsize + 
1pt,black,opacity=.04,rounded corners=\boxroundness] 
(\tmpa, \tmpa) rectangle +(\boxw - \tmpa - \tmpa, \boxh - \tmpa - 
\tmpa);
}

\filldraw[fill=white!50, draw=black!80, rounded corners=\boxroundness] (0, 
0) rectangle (\boxw, \boxh);
\draw node[xshift=\boxroundness,yshift=\boxroundness,inner sep=0pt,outer 
sep=0pt,anchor=south west] (0,0) {\usebox{\shadowblockbox}};
\end{tikzpicture}\vspace{0cm}%
}
\newcommand{\finalmisuses}{\checknum{49}\xspace} %% SO only
\newcommand{\somisuses}{\checknum{28}\xspace}
\newcommand{\ghmisuses}{\checknum{21}\xspace}

\newcommand{\totalghcommits}{\checknum{358}\xspace}

\newcommand{\distinctAPIs}{\checknum{45}\xspace}

\newcommand{\methodPerc}{\checknum{43\%}\xspace}
\newcommand{\paramPerc}{\checknum{51\%}\xspace}
\newcommand{\conditionPerc}{\checknum{6\%}\xspace}

\newcommand{\doccaveatsPerc}{\checknum{39\%}\xspace}

\newcommand{\datadep}{\checknum{27}\xspace}
\newcommand{\datadepPerc}{\checknum{55\%}\xspace}

\newcommand{\lib}{\checknum{5}\xspace}

\newcommand{\misuseWithIncorrectOutput}{\checknum{17}\xspace}
\newcommand{\misuseWithIncorrectOutputPerc}{\checknum{35\%}\xspace}

\newcommand{\misuseWithErrorsPerc}{\checknum{41\%}\xspace}

\newcommand{\misuseWithMemoryPerc}{\checknum{20\%}\xspace}

\newcommand\checknum[1]{\textcolor{black}{#1}}

\newcommand{\dsml}{data-centric}
\newcommand{\Dsml}{Data-centric}

\newcommand{\cav}{API directive}
\newcommand{\cavs}{API directives}
\newcommand{\so}{Stack Overflow}

\newcommand{\etal}{et al.}

\usepackage{amsthm}
\theoremstyle{definition}
\newtheorem{defn}{Definition}
\begin{document}

\title{An Empirical Study of API Misuses of Data-Centric Libraries}

\author{Akalanka Galappaththi}
\email{akalanka@ualberta.ca}
\orcid{https://orcid.org/0000-0002-6756-6610}
\affiliation{%
  \institution{University of Alberta}
  \city{Edmonton}
  \country{Canada}
}

\author{Sarah Nadi}
\email{sarah.nadi@nyu.edu}
\orcid{https://orcid.org/0000-0002-0091-6030}
\affiliation{%
  \institution{New York University Abu Dhabi}
  \city{Abu Dhabi}
  \country{United Arab Emirates}
}

\author{Christoph Treude}
\email{ctreude@smu.edu.sg}
\orcid{https://orcid.org/0000-0002-6919-2149}
\affiliation{%
	\institution{Singapore Management University}
	\country{Singapore}
}

\renewcommand{\shortauthors}{Galappaththi et al.}

\begin{abstract}

Developers rely on third-party library Application Programming Interfaces (APIs) when developing software. However, libraries typically come with assumptions and API usage constraints, whose violation results in \textit{API misuse}. API misuses may result in crashes or incorrect behavior. 
Even though API misuse is a well-studied area, a recent study of API misuse of deep learning libraries showed that the nature of these misuses and their symptoms are different from misuses of traditional libraries, and as a result highlighted potential shortcomings of current misuse detection tools. 
We speculate that these observations may not be limited to deep learning API misuses but may stem from the data-centric nature of these APIs.
Data-centric libraries often deal with diverse data structures, intricate processing workflows, and a multitude of parameters, which can make them inherently more challenging to use correctly. 
Therefore, understanding the potential misuses of these libraries is important to avoid unexpected application behavior.  
To this end, this paper contributes an empirical study of API misuses of five data-centric libraries that cover areas such as data processing, numerical computation, machine learning, and visualization. 
We identify misuses of these libraries by analyzing data from both Stack Overflow and GitHub. 
Our results show that many of the characteristics of API misuses observed for deep learning libraries extend to misuses of the data-centric library APIs we study. 
We also find that developers tend to misuse APIs from data-centric libraries, regardless of whether the \cav{} appears in the documentation. 
Overall, our work exposes the challenges of API misuse in data-centric libraries, rather than only focusing on deep learning libraries. Our collected misuses and their characterization lay groundwork for future research to help reduce misuses of these libraries.

\end{abstract}

\begin{CCSXML}
  <ccs2012>
     <concept>
         <concept_id>10011007.10011074.10011111.10011696</concept_id>
         <concept_desc>Software and its engineering~Maintaining software</concept_desc>
         <concept_significance>500</concept_significance>
         </concept>
   </ccs2012>
\end{CCSXML}
  
\ccsdesc[500]{Software and its engineering~Maintaining software}

\keywords{API misuse, data-centric libraries, empirical study}

\maketitle

\section{Introduction}
\label{sec:intro}

When developing software applications, developers typically use third-party libraries that offer access to various functionality through a set of \textit{Application Programming Interfaces} (APIs).
%For example, developers may use a library that gives them access to a \texttt{plot()} function to create a graph.
While some APIs are easy to use and integrate, others have certain usage constraints that application developers need to follow in order to correctly achieve the desired functionality. Violating these constraints leads to incorrect API usage, also referred to  as an \emph{API misuse} \cite{Detector-Eval-Amann:2019, FUM:2022, API-graph-Zeng:2021}.
API misuses can lead to program crashes, security vulnerabilities, performance problems, or unexpected program behavior~\cite{FUM:2022}. 

There is a long line of research studying API misuses \cite{PR-Miner-Li:2005, Jaddet-Wasylkowski:2007, GrouMiner-Nguyen:2009, Alattin-Thummalapenta:2009, MuDetect-Amann:2019, MisuseHint-Liang:2022, API-KG-Ren:2020, API-graph-Zeng:2021, MutAPI-Wen:2019}, most of which focused on Java APIs and typically considered misuses related to the control flow and data flow between APIs~\cite{FUM:2022,MuDetect-Amann:2019}.
For example, a common API misuse is missing \texttt{null} checks or forgetting certain API calls (e.g., not calling \texttt{close()} after calling \texttt{open()} on a stream).

In recent work, Wei \etal{} \cite{dl-api-misuse-llm:2024} investigated API misuses of two Python deep learning libraries, TensorFlow and PyTorch. 
They found that many of the misuses are caused by incorrect device usage (e.g., CPU instead of GPU) or under-the-hood data conversion problems. 
They conclude that misuses like data shape mismatch in deep learning libraries are unique, because they are harder to detect, and also because they may not immediately raise errors but instead produce incorrect results or lead to performance problems. %Developers need to have domain expertise and carefully inspect the source code to identify such misuses, because these bugs are often tied to the nature and structure of data being processed.
However, we observe that apart from CPU/GPU misuses, many of their observed misuses may not be specific to deep learning per se but may rather stem from the reliance on data.
We also observe that there are many other libraries that focus on processing, analyzing, and deriving insights from data.
For example, data processing and manipulation libraries such as pandas \cite{pd-lib:23} or visualization libraries such as seaborn~\cite{sb-lib:23} all deal with diverse data structures, intricate processing workflows, and a multitude of parameters, which can make them inherently more challenging to use correctly.
We use the term \textit{data-centric libraries} to refer to such libraries.
Given their shared focus on handling data, it is reasonable to speculate that some of the new types of API misuses observed for deep learning libraries may also manifest in other \dsml{} libraries.

Consider the following deep learning API misuse example provided by Wei et al. \cite{dl-api-misuse-llm:2024}: a developer intends to multiply two tensors, $A$ and $B$, where $A$ is a $ 3 \times 3$ tensor and $B$ is a $2 \times 3$ tensor. Since the dimensions are not compatible for multiplication, the developer needs to first transform tensor $B$.
In TensorFlow, when transforming a $ 2 \times 3 $ tensor to a $3 \times 2$ tensor, the developer has two options: \texttt{reshape} or \texttt{transpose}. While both methods produce tensors with similar shapes, \texttt{transpose} is the correct API to call in this scenario, because with \texttt{transpose}, the dimensions are appropriately matched ($[3\times3] \times [3\times2] \rightarrow [3\times2]$) and elements are correctly positioned. While using \texttt{reshape} may not immediately appear incorrect since it returns expected dimensions, using this API leads to incorrect output as the elements of $B$ are rearranged improperly.
Importantly, this misuse does not trigger an error but rather yields an incorrect output. Any subsequent operations using the resulting tensors are also compromised.

\begin{figure}[t!]
	\begin{minipage}{0.5\textwidth}
		\begin{subfigure}{\textwidth}
			
\begin{lstlisting}[style=pythoncode]
import pandas as pd
import matplotlib.pyplot as plt
import seaborn as sns

# df dataframe has format:
#    x_values  y_values  color
# 0  0.200000  0.333333    2.0
# 1  0.003334  0.234043    1.0
# .. ..	       ..	   ..
# 7  0.003334  0.234043    1.0

df = ...

          #violet     #green     #orange
colors = ['#747FE3', '#8EE35D', '#E37346']
- sns.set_palette(colors)

- sns.scatterplot(data = df,  x="x_values", y="y_values", hue="color", s=40, legend=False)
+ sns.scatterplot(data = df,  x="x_values", y="y_values", hue="color", s=40, legend=False, palette=colors)
plt.show()
\end{lstlisting}
			\caption{Seaborn API misuse of \texttt{set\_palette} and \texttt{scatterplot}}
			\label{fig:seabornmisuse}
		\end{subfigure}
	\end{minipage}
	\hfill
	\begin{minipage}{0.5\textwidth}		
		\begin{subfigure}[t]{0.45\textwidth}
			\centering
			\includegraphics[width=\textwidth]{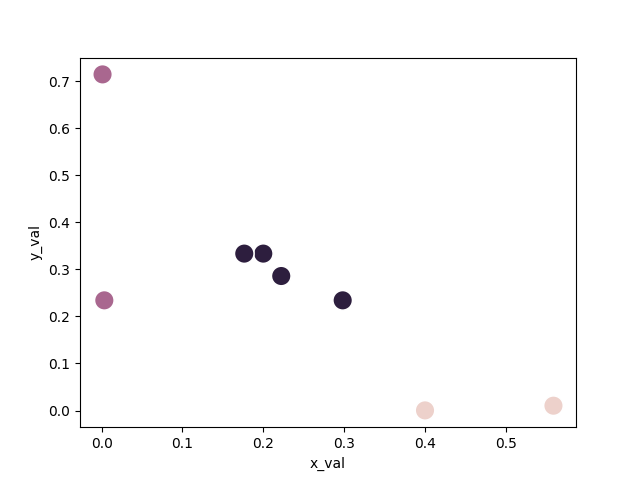}
			\caption{Incorrect hue colors}
			\label{fig:seaborn-incorrectgraph}
		\end{subfigure}
		% \vfill
		\begin{subfigure}[t]{0.45\textwidth}
			\centering
			\includegraphics[width=\textwidth]{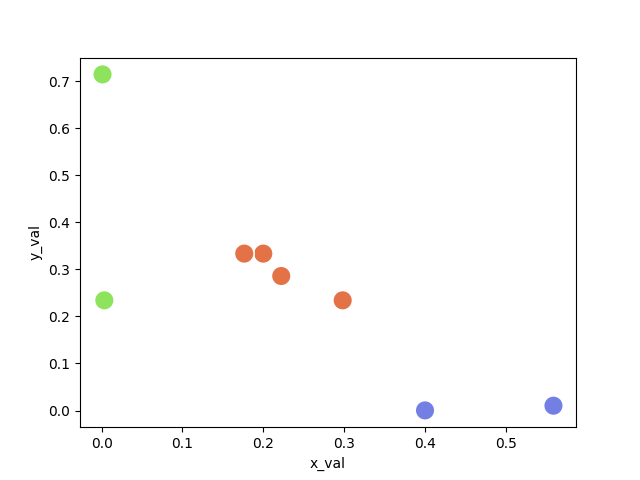}
			\caption{Correct hue colors}
			\label{fig:seaborn-correctgraph}
		\end{subfigure}
	\end{minipage}
%	\hfill
	\vspace{-0.3cm}
	\caption{Example of seaborn API misuse and its impact, based on Stack Overflow question 67637829.\vspace{-0.6cm}}
	\label{fig:seaborn-example}
\end{figure}

Now, let us consider the scenario shown in Figure~\ref{fig:seabornmisuse} where a developer 
uses the Python visualization library seaborn to create a scatterplot. 
In this example, the developer wants to color points on the scatterplot differently according to the value in the column \texttt{color} of their data frame, which is why they set \texttt{hue="color"} on Line 18.
The variable \texttt{colors} on Line 15 contains the specific colors they want to use.
Seaborn offers different options to set color palettes such as calling the \texttt{set\_palette(...)} method or passing a value to the \texttt{palette} parameter in the \texttt{scatterplot} API (e.g., \texttt{scatterplot(..., palette=...)}). 
On Line 16, they use \texttt{set\_palette} to assign these colors as the palette to use.  
However, the displayed scatter plot in Figure~\ref{fig:seaborn-incorrectgraph} colors the points using a different palette than the one specified; the correctly colored scatterplot is shown in Figure~\ref{fig:seaborn-correctgraph}.
It turns out that the use of the specified color palette depends on the type of data in the column used as the \texttt{hue}.
In cases of numerical values, seaborn defaults to an internal color palette, completely ignoring whatever value was set using \texttt{set\_palette}.
Had the type of the \texttt{color} column been categorical (i.e., strings) such as ``zero'', ``one'', ``two'', the same code would have correctly used the specified colors.
The only way to get the desired behavior with a numerical column is to pass \texttt{palette=colors} as an argument to \texttt{scatterplot} on Line 19 instead of calling \texttt{set\_palette} on Line 16.
Thus, the data inside the data frame affects the correctness of the API usage, highlighting the importance of understanding how data types influence the behavior of the library. 
% This example underscores the potential for confusion and the need for careful debugging, even though the impact may seem low (incorrect colors).

These examples demonstrate that some of the challenges and unique nature of API misuses found for the two deep learning libraries Tensorflow and PyTorch~\cite{dl-api-misuse-llm:2024, gDefects4DL-icse:22, TADAF:2022} may extend to other commonly used Python libraries, specifically those with a \mbox{\dsml{}} nature. In general, misuses caused by data conversion errors are not necessarily limited to deep learning or even machine learning libraries.
Our goal in this work is to more broadly investigate API misuses of \dsml{} libraries.

%Data has generally become a key ``ingredient'' in many modern software applications, making \dd{} application development~\cite{HeseniusDD2019} more pervasive. 
%Python is currently the most popular programming language for developing such \dd{} appplications~\cite{python_dsml_popularity:2008, kaggle-survey:2022}, offering many libraries for processing data to help develop these applications.
%While previous research has mostly concentrated on exploring bugs and API misuses when using deep learning libraries~\cite{TF-bugs-Zhang:2018,DL-bug-char-Islam:2019,dl-api-misuse-llm:2024}, we argue that there are other Python libraries that are instrumental in processing, analyzing, and deriving insights from vast amounts of data.

% Specifically, this paper presents an empirical study of API misuse in the scope of \dsml{} libraries. 
% We focus on Python libraries since Python is currently the most popular programming language for developing \dd{} applications~\cite{python_dsml_popularity:2008, kaggle-survey:2022}. 
% We select five popular \dsml{} Python libraries~\cite{popular-ds-libs:2020}: NumPy \cite{np-lib:23}, pandas \cite{pd-lib:23}, scikit-learn \cite{sk-lib:23}, Matplotlib \cite{mp-lib:23}, and seaborn \cite{sb-lib:23}. Note that deep learning libraries are also data-centric libraries, but we exclude them from our study as our goal is to investigate that misuses identified in them are beyond the deep learning domain.

Specifically, this paper presents an empirical study of API misuse in data-centric libraries. We focus on Python, the most popular language for data-driven applications~\cite{python_dsml_popularity:2008, kaggle-survey:2022}. Since deep learning libraries have been already studied~\cite{dl-api-misuse-llm:2024}, we selected five additional widely-used \dsml{} Python libraries: NumPy~\cite{np-lib:23}, pandas~\cite{pd-lib:23}, scikit-learn~\cite{sk-lib:23}, Matplotlib~\cite{mp-lib:23}, and seaborn~\cite{sb-lib:23}. 
%We study API misuses of these libraries by mining data from both Stack Overflow and GitHub, and verifying the misuses through the library's documentation.% and the creation of minimal reproducible code examples.

To identify misuses, we manually analyze 345 Stack Overflow posts and \totalghcommits commits from open-source GitHub repositories 
 that use these libraries. 
%Leveraging information in Stack Overflow posts, including the provided code examples, problem description, and stack traces or output, we determine if a post is describing a problem rooted from an API misuse. 
%Similar to previous work that used git history to discover API misuses through the changes developers make to their code \cite{Mubench-Amann:2016,dl-api-misuse-llm:2024,misuse-wild-Li:2021}, we also manually analyze \totalghcommits commits from open-source GitHub repositories that use the five selected libraries.
Using information from these two different data sources allows us to discover both API misuses that make their way to developer's committed code, as well as those where developers seek community help before finalizing their code.
We then categorize the identified misuses in terms of the misuse type, exhibited symptom, and root cause using Wei \etal's taxonomy \cite{dl-api-misuse-llm:2024}, which is derived from the study of misuses of deep learning libraries.
We additionally investigate whether the library documentation includes any explicit guidelines for avoiding the identified misuses. 
% The presence or absence of such guidelines provide valuable insights into the underlying API misuses, shedding light on the assumptions made by library developers, the level of background knowledge expected from API users, and the overall usability of the API \cite{API-hard-learn-Robillard:2009, usability-Myers:2016}.

% We manually investigate \todo{345} Stack Overflow posts related to these libraries to identify API misuses. Then we manually investigate \todo{x} commits from 100 GitHub projects to search misuse instances identified by analyzing Stack Overflow.

%Given the varying definitions of what an API misuse is (discussed in detail in Section~\ref{defs}), we develop a clear coding guideline to define the scope of what we consider as a misuse.
% For each confirmed misuse, we also note its impact, the \cav{} being violated, and whether the \cav{} is explicitly documented.\sn{perhaps we should be more explicit about what we do that is previous established methods (i.e., categorize misuse and its impact) vs. what we do that is new, regardless of the novelty of the target libraries}
% An \textit{\cav{}} is a guideline that describes how to correctly use an API~\cite{FUM:2022}.
% For further verification, we develop a minimal reproducible example for each misuse to ensure that our understanding of the misuse and the required fix is correct.
% We categorize types of misuses and their impact using the classification labels generated by Wei \etal~\cite{dl-api-misuse-llm:2024}.

Overall, we collect \finalmisuses API misuses, covering \distinctAPIs{} distinct APIs.
Our findings reveal that despite differences in programming paradigms between deep learning and traditional machine learning \cite{dl-api-misuse-llm:2024} or visualization libraries, the nature of misuses previously observed extend beyond the deep learning libraries.
Specifically, we find that \checknum{39\%} of the misuses are due to data-conversion errors and the most misused element is API parameters (\checknum{51\%}).
However, we find the need to extend the existing taxonomy with an additional dimension of \textit{data dependency} to capture the idea that the exact same API usage may be a misuse in one case while a correct usage in another, only depending on the data being processed.
The example discussed in Figure~\ref{fig:seabornmisuse} is an example of a data-dependent misuse.
We find that \datadepPerc of the studied misuses are data dependent and \misuseWithIncorrectOutputPerc of misuses result in incorrect output without any explicit run-time errors, while \checknum{41\%} cause program crashes.
% The majority of the misuses (\checknum{41\%}) caused program crashes, which indicates the severity of the \dsml{} misuses.
Overall, we find that the common characteristic between all these libraries that leads to the API misuses is their heavy reliance on data. 
Surprisingly, we find that \doccaveatsPerc of the misuses have documented \cavs, but developers still misused the API.

% and that only \misuseWithErrorsAndUsefulMsgPerc of the run-time errors \sn{this isn't discussed in the results.. Decide what will be kept/removed and make sure to streamline everything} for the remaining misuses had information that could be used to identify the misused API. %Overall, our study shows that developers tend to misuse APIs from data-centric libraries regardless of the availability of documented \cavs. \sn{not really the main conclusion you want to end with, and also already mentioned in the first sentence?}

Our findings have implications for language and API design, misuse detectors, and surfacing information buried in documentation.
Specifically, \dsml{} libraries need to make assumptions about data content and format and yet there are no built-in language mechanisms for helping API designers enforce these assumptions. Even when they document these assumptions, client developers end up misusing them, further implying that we need more research on surfacing important information in documentation. 
%While third-party library APIs typically include checks for data format (e.g., pandas dataframe or NumPy array), ensuring the integrity of data content, such as detecting mixed data types within a column of a pandas dataframe, poses a significant challenge in API design.
\newpage
In summary, this paper makes the following contributions:
\begin{itemize}[leftmargin=*,topsep=0pt]
	\item We define the notion of data-centric libraries, and show that API misuses of those libraries share characteristics with previously identified API misuses of deep learning libraries.
	\item We collect \finalmisuses API misuses from \lib Python libraries, corresponding to misuses of \distinctAPIs distinct APIs.
	\item We categorize the detected misuses using an existing deep-learning misuse taxonomy, finding that it is expressive enough to categorize the majority of the data-centric misuses. However, we had to introduce the notion of data dependency to the taxonomy to capture some of our observations.
	\item We identify if the misused APIs have corresponding guidelines in the library documentation.% that could have helped prevent the misuse.
	\item For each misuse, we construct a reproducible example to illustrate the misuse. %, and provide the associated \cav, whether the \cav{} is documented, and the impact of the misuse.
	
	\item We discuss the implications of our findings on API and language design, surfacing information buried in documentation, and designing misuse detectors.
\end{itemize}

Our data is available on our online artifact page \cite{dataset}.

\section{Background, Scope, and Definitions}
\label{sec:background}

\subsection{Definitions and Scope}

Schlichtig et al.~\cite{FUM:2022} noted some discrepancies in how researchers define API misuse. 
For example, some authors consider only the external API of a library~\cite{Detector-Eval-Amann:2019, misuse-wild-Li:2021}, while others consider also internally defined APIs that are meant to be used only inside the current application \cite{dynamic-misuse-He:2023}.
Similarly, some work considers client code that uses an outdated API due to breaking changes in the library as a misuse \cite{dynamic-misuse-He:2023, FUM:2022, dl-api-misuse-llm:2024}, while other work considers general Python typing issues as misuses \cite{dynamic-misuse-He:2023}.
In this work, we are interested in API misuses of third-party libraries that stem from the unique nature of the \dsml{} domain, rather than the nature of the underlying programming language or general software evolution characteristics.
To clearly define our scope, we use the following definitions from Schlichtig et al.~\cite{FUM:2022}, with some adaptations, if needed, shown in non-italic square parentheses.
%This focused scope allows us to isolate and understand the unique challenges associated with the misuse of data-centric library APIs.   

% \vspace{-0.2cm}
\begin{defn}\label{def:api}
    An \textit{\textbf{Application Programming Interface} (API)} is the \textit{``public interface {\normalfont [that]} exposes software elements (e.g., classes and methods) to the outside world, making the implemented functionality accessible~\cite{FUM:2022}.''} {\normalfont We focus only on the use of third-party library APIs in client code, considering all public API elements such as classes, methods (including their parameters), and attributes.}
\end{defn} 
 
% The \texttt{scatterplot} method in Figure~\ref{fig:seabornmisuse} is an API from the \texttt{seaborn} library.
% \vspace{-0.2cm}
\begin{defn}\label{def:directive}
    ``\textit{An \textbf{API directive} is a natural-language statement related to guidelines or constraints that describes how to use an API correctly and optimally {\normalfont [, regardless of the developer's task or intention]}. It can be part of the underlying documentation of an API. However, an API directive can also be implicit, for example, because of incomplete documentation or expected domain-specific knowledge}~\cite{FUM:2022}.''
\end{defn}

% \vspace{-0.2cm}
\begin{defn}\label{def:usageconstraint}
    ``\textit{An \textbf{API usage constraint} is an API directive
that restricts the actual use of an API. These restrictions are not enforced by the programming language itself, such as correct typing {\normalfont[, nor are they part of the natural software evolution process (e.g., deprecation)]}. Because API usage constraints are API directives, they are imposed by the API designer/expert}~\cite{FUM:2022}.'' 
\end{defn}

For example, the \texttt{scatterplot} documentation contains this \cav{}, related to the misuse in Figure~\ref{fig:seaborn-example}: \textit{``The default treatment of the hue (and to a lesser extent, size) semantic, if present, depends on whether the variable is inferred to represent “numeric” or “categorical” data. In particular, numeric variables are represented with a sequential colormap by default, [...] This behavior can be controlled through various parameters, as described [...] below.~\cite{sb_scat:23}''}.
%\Cavs{} can be explicitly present in the documentation, or can be implicit without any documented information.

% \vspace{-0.2cm}
\begin{defn}\label{def:misuse}
    ``\textit{An \textbf{API misuse} is the violation of one or more API usage constraints. Such violation leads to misbehaviour of the API, e.g. errors, crashes, {\normalfont [incorrect output,]} or vulnerabilities}~\cite{FUM:2022}.'' 
\end{defn}

%% ag{I commented this out, bacuse ICSE paper also talk about incorrect output a significant problem.}
% The example from Figure~\ref{fig:seaborn-example} demonstrates that not all misuses result in an (eventual) crash of the program or even an (Type) error in Python; in this specific case, there was a ``silent'' failure where the produced output was incorrect.

% \vspace{0.1cm}
Note that the API misuse definition depends on the definition of an API usage constraint.
The distinction in Definition \ref{def:misuse} is important for our work, since we do not want to consider errors that are inherent to programming in Python regardless of the library being used.
For example, if a developer passes a float argument to an integer parameter, the Python type system will raise a run-time TypeError \textit{integer argument expected, got float}.
%consider the NumPy \texttt{insert(...)}~\cite{np-insert} function, which expects an integer value for the optional parameter \texttt{axis}. 
%If the developer passes a float instead, the Python type system will raise a run-time TypeError \textit{integer argument expected, got float}.
%While this is an incorrect usage of a third-party library API, 
We do \textit{not} consider this as an API misuse since it is a typical programming error that stems from the dynamic typing nature of Python rather than violating an API usage constraint.
%For example, consider the \texttt{scatterplot} function from Figure~\ref{fig:seabornmisuse}.
%The function expects an integer value for the parameter \texttt{s}. 
%If the developer passes a string instead, the Python type system will raise a run-time ValueError \textit{s must be a scalar, or float array-like with the same size as x and y}.
%While this is an incorrect usage of a third-party library API, this is \textit{not} an API misuse since it is a typical programming error that stems from the dynamic typing nature of Python rather than violating an API usage constraint as defined above.
This is, for example, a key distinction between our definition of API misuse (as well as Wei \etal~\cite{dl-api-misuse-llm:2024}'s) and that of He et al.~\cite{dynamic-misuse-He:2023}'s study of general Python API misuses. 

Our explicit exclusion of deprecations from Definition \ref{def:misuse} is also important. 
Deprecations and breaking changes are a general phenomenon for all dependencies/libraries, regardless or the programming language or nature of these libraries, and there are typically warnings in place for such deprecations. 
%Additionally, when a library plans to introduce changes to an existing API in future a version (e.g., replacing a function), the Python interpreter typically provides warnings of future deprecation of the API. 
Thus, we do not consider modifications to an API usage that occur due to compliance with future library changes, as such adaptations are inevitable with the evolution of libraries. While Wei et al.'s study \cite{dl-api-misuse-llm:2024} considered deprecation management errors as misuses, accounting for approximately 20\% of the collected misuses, deprecations are not specific to deep learning libraries in any way. Our focus is on identifying misuses in the domain of data-centric libraries, while deprecations would be observed for any library from any domain. %Moreover, modern IDEs often notify developers about deprecated APIs, enabling them to address such issues before code reaches the production environment.

Note the slight variation we added in Definition \ref{def:usageconstraint} of API directive, w.r.t. the developer's intention.
For example, a developer accidentally passing ``write'' as the open mode of a file instead of ``append'' is not a misuse since both these modes are correct and depend on what the developer wants to do. We cannot assume the developer's intention of whether they want to write or append to the file.
A misuse should always be a misuse, without requiring extra information about the developer's intention.
In contrast, setting the axis color of a plot without enabling the axis first is a misuse since any code that sets the axis color without enabling the axis is problematic (the axis color would not have any effect).

Finally, we note one interesting distinction by Schlichtig et al.~\cite{FUM:2022} in the following definition:

% \vspace{-0.2cm}
\begin{defn}\label{def:dom-api}
    \textit{``A \textbf{domain-specific API} offers functionality tailored to a specific domain. Its domain determines the achievable goal and application rules of a domain-specific API. In contrast, \textbf{non-specific APIs} are not tailored to a domain, nor do they determine a specific goal associated with their use~\cite{FUM:2022}.''} 
\end{defn}
% \vspace{-0.4cm}

% \vspace{0.2cm}
Schlichtig et al.~\cite{FUM:2022} argued that this distinction is important because domain-specific APIs sometimes have additional or very specific types of usage constraints, as exhibited by API misuse in the cryptography domain for example~\cite{Cognicrypt-Kruger:2017,NadiCryptoAPIs16}.
\Dsml{} libraries offer domain-specific APIs, and previous studies of deep learning libraries already showed that characteristics such as accounting for data shape in these libraries caused new types of misuses, posing challenges to traditional misuse detection tools.
Both of these facts motivate the study in this paper, where we extend the scope beyond only deep learning libraries to understand API misuses in the general domain of \dsml{} libraries.

\vspace{-0.3cm}
\subsection{API Misuse of Deep Learning Libraries}
\label{sec:background-taxonomy}

%Wei \etal's taxonomy amalgamates various categories from prior studies, such as Li \etal's violation types \cite{misuse-wild-Li:2021} and some of Amann \etal's API element types \cite{MuDetect-Amann:2019}.

\begin{figure}[t!]
	\includegraphics[scale=0.32]{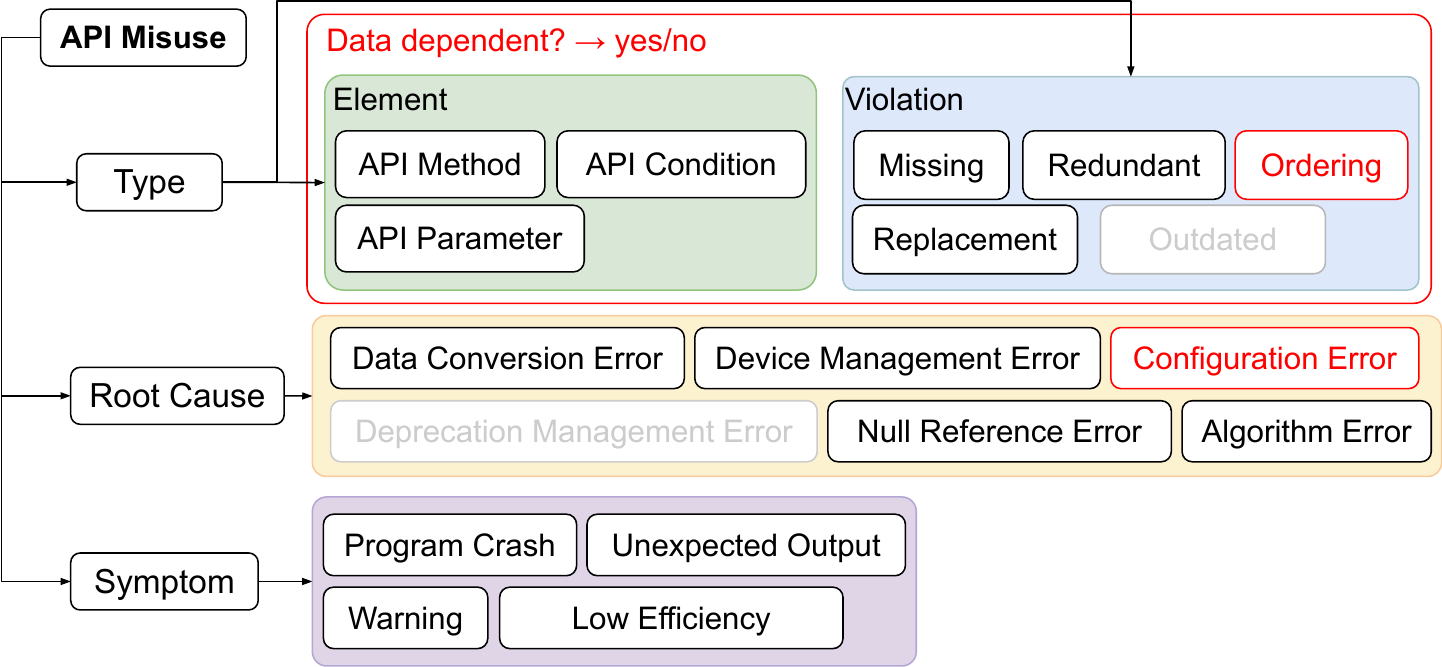}
    \vspace{-0.3cm}
	\caption{Our updated misuse classification taxonomy, based on Wei et al.~\cite{dl-api-misuse-llm:2024}'s analysis of deep learning misuses. Greyed out boxes are out-of-scope items we do not consider, while red boxes show our additions.}
    \vspace{-0.5cm}
	\label{fig:taxonomy}
\end{figure}

Our study is motivated by the findings of Wei et al.~\cite{dl-api-misuse-llm:2024} who examined API misuse of Python deep learning libraries, particularly TensorFlow and PyTorch. 
To identify misuses, they mined the change history of various client repositories of these two libraries and manually analyzed candidate commits that have made changes to the libraries' API usages.
Based on the identified misuses, they create a taxonomy of API misuse types, root causes, and impact.
While this taxonomy amalgamates information from existing work~\cite{dynamic-misuse-He:2023,FUM:2022,Detector-Eval-Amann:2019,misuse-wild-Li:2021}, the authors add new root causes that stem from the nature of deep learning libraries, specifically data conversion problems and device management errors.
Figure~\ref{fig:taxonomy} shows the summary of their developed taxonomy, which we leverage for labeling the misuses we discover in our work.
The taxonomy first categorizes the types of misuses by the program element involved in the misuse as well as the type of the violation (e.g., misuse in Figure~\ref{fig:seaborn-example} would be a missing API parameter). API method refers to misuses related to calling functions or methods. API parameter are the parameters of the API methods. API condition refers to conditional statements required for the API usage.
%Violation types missing, redundant, replacement, and deprecation are self-explantory. The violation type \textit{ordering} refers to the incorrect order of API calls.
The taxonomy also categorizes the root cause of the problem. 
Devise management errors relate to any hardware or resource utilization (e.g., using CPU vs GPU).
Algorithm errors are related to mathematical problems (e.g., dividing by zero) or incorrect calculations. Data conversion errors represent issues that relate to incorrect conversions between data types or shapes. When a program accesses a null object, null reference errors occur. Deprecation management errors relate to the usage of deprecated APIs or parameters. 
%In our same example, the root cause would be a \textit{state handling error}. This category is not in the taxonomy as such types are rarely found in deep learning misuses. 
Finally, the taxonomy categorizes the observed symptom, which in our example would be unexpected output. Program crashes and warnings refer to runtime errors and warnings. Low efficiency refers to slow program execution. 
% \sn{FYI: we should not discuss new results in the background so I removed the statement about order and state handling errors. The note in the next paragraph should suffice}

Note that the figure shows some of our modifications to the taxonomy in terms of adding items (in red, discussed later in our results), or not considering items due to our goal and defined scope above.
For the latter, we do not consider the greyed out outdated violation and root cause of deprecation management errors.

\vspace{-0.3cm}
\section{Methods}

We conduct an empirical study to identify API misuses in \dsml{} libraries.
Most previous research on API misuse identified misuses from either \so{} or git history on GitHub~\cite{DL-bug-char-Islam:2019, DL-taxonomy-Humbatova:2020, TF-bugs-Zhang:2018}.
In our work, we use both data sources to collect API misuses.
We manually analyze \so{} questions and their answers to determine if they are discussing a problem related to API misuse. We also investigate commits from Python projects on GitHub, as they may contain fixes for API misuses. For confirmed misuses from either data source, we note how the misuse happened (e.g., passing incorrect argument type) and what was the observed symptom (e.g., incorrect output) by reproducing the misuse. We then use Wei \etal~\cite{dl-api-misuse-llm:2024}'s misuse classification of deep learning API misuses  to categorize the misuse types, their root causes, and symptoms. As a new contribution of our work, we consult the library’s documentation to determine if there is a documented \cav{} related to the misuse. 

\vspace{-0.3cm}
\subsection{Data Collection}
% \subsection{Analyze Stack Overflow Posts to Identify Misuses}

\vspace{-0.2cm}
\subsubsection{Libraries}  We select five popular \dsml{} Python libraries \cite{popular-ds-libs:2020}: NumPy \cite{np-lib:23}, pandas \cite{pd-lib:23}, scikit-learn \cite{sk-lib:23}, Matplotlib \cite{mp-lib:23}, and seaborn \cite{sb-lib:23}. NumPy is a fundamental package for scientific computing with Python, providing support for large, multidimensional arrays and matrices, along with a collection of mathematical functions to operate on these arrays. Pandas is a fast, powerful, flexible, and easy-to-use data analysis and data manipulation library. Scikit-learn is a popular library for data preprocessing, machine learning, and visualization. Matplotlib and seaborn are comprehensive libraries for creating interactive visualizations in Python. We consider both Matplotlib and seaborn as they are both commonly used together for visualizations.

\vspace{-0.2cm}
\subsubsection{Selecting questions in Stack Overflow}
\label{subsec:so-data}

\so{}~\cite{se-survey:2023} is a widely used question and answer website that is popular among developers and programming enthusiasts. When posting questions, users are encouraged to provide minimal reproducible code examples, stack trace of errors if any, and problem descriptions. \so{} has tags that represent various technologies, which can highlight the libraries discussed in questions. We utilize these characteristics in \so{} questions to identify questions that potentially relate to API misuses.

%Our goal is to identify misuses of the selected five data-centric libraries.
Since \so{} has tags to represent various libraries, we retrieve \so{} posts tagged with each of our libraries using the Stack Exchange data explorer \cite{sede}. To increase our chances of finding questions related to API misuse, we use the following additional filtering criteria. We remove questions with no answers since answers are essential to help us determine the root cause behind the posed problem. We filter out questions whose title starts with ``how to'' since these relate to the poster asking about how to implement particular functionality, which is unlikely to relate to API misuse. We also filter out questions with negative scores, because a negative score indicates poorly articulated questions or those missing information~\cite{ess-Nadi:2020}. Additionally, we only collect the questions that were posted between 2019 and 2023 in order to avoid analyzing outdated questions. 
%% \ag{Does not seems add much value}
% Note that some of the visualization questions are tagged with both Matplotlib and pandas as they use data from pandas dataframes for visualization. To avoid having questions duplicated between our Matplotlib and pandas data, we filter out questions tagged with Matplotlib when collecting questions for pandas. 
Finally, we filter out questions with no code snippet in the question body, as we need to analyze a code snippet to determine whether it contains an API misuse. 

Table \ref{tab:so_q_stat} shows the total number of questions in \so{} tagged with each library and the number of questions that we collect after applying the above filtering criteria.
Since manually analyzing close to 164,000 posts is infeasible, we resort to analyzing a sample of these threads. 
%When analyzing the distribution of scores of these questions, 
We find that the question scores exhibit a right-skewed distribution where the majority of the questions either have a score of zero, one, or two while only a few questions have a score higher than two. Accordingly, we partition the questions into three groups based on score: zero, one or two, and three or higher and then we randomly select questions from each group to annotate (i.e., we perform stratified sampling according to question score). We choose 90\% confidence level and 10\% error margin when selecting samples from each question group (i.e. tag). The resulting sample sizes varied between 67-69 per tag, which provided us with a target of 340 questions to be analyzed. For the sake of dividing questions equally among three annotators, we analyze 345 \so{} questions in total.

\begin{table}[t!]
	\centering
	\caption{Number of collected and filtered Stack Overflow questions for each library, collected on 31-05-2023.\vspace{-0.3cm}}
	\resizebox{0.42\textwidth}{!}{
	\begin{tabular}{lrrrr}
		\toprule
		\multicolumn{1}{l}{\textbf{Tag}} & 
		\multicolumn{1}{r}{\textbf{Total \# of}} &
		\multicolumn{1}{r}{\textbf{\# of questions}} &
		\multicolumn{1}{r}{\textbf{Minimum}} &
		\multicolumn{1}{r}{\textbf{Annotated}}\\
		& 
		\multicolumn{1}{r}{\textbf{questions}} &
		\multicolumn{1}{r}{\textbf{after filtering}} & 
		\multicolumn{1}{r}{\textbf{sample size}} &
		\multicolumn{1}{r}{\textbf{sample size}}
		\\
		\midrule
		pandas & 106,331 & 95,992 & 69 & 69\\
		NumPy & 30,698 & 27,774 & 68 & 69\\
		scikit-learn & 7,132 & 6121 & 68 & 69\\
%		TensorFlow & 24,597& 20,105 & 68 & 69\\
		Matplotlib & 16,916 & 15,463 & 68 & 69\\
		seaborn & 3,251 &  2,942 & 67 & 69\\
		
		\midrule
		Total & 164,328 & 148,116 & 340 & 345\\
		\bottomrule	
%		\multicolumn{2}{l}{\footnotesize*Number of questions after applying filters in Stack Exchange data explorer.}\\
		
	\end{tabular}
	}
	\label{tab:so_q_stat}
	\vspace{-0.5cm}
\end{table}

\vspace{-0.2cm}
\subsubsection{Selecting commits in GitHub projects}
\label{subsec:gh-data}
%GitHub contains over 35 million Python projects. 
Our goal is to find Python projects that use our libraries of interest and analyze their commit history to identify potential fixes of API misuses of these libraries.
%However, not all of these projects use the libraries that we selected for this study. Unlike \so{} tags, GitHub does not provide a direct way to identify the dependencies of Python projects. While Python projects commonly use requirements.txt file or setup.py file to declare dependencies \cite{python-dep:23}, some projects do not contain those files. Thus, we consider an alternative approach, that is to 
For each analyzed library, we use the GitHub dependency graph of the library's public repository to find client projects.
%We scrape the web pages in each library’s dependents to find repositories that use the library. 
% We collect 813,074 dependent projects for pandas, 846,891 for NumPy, 411,242 for scikit-learn, 647,415 for Matplotlib, and 205,566 for seaborn.
%Table \ref{tab:gh_dep_stat} shows the number of dependent projects for each library. 
After collecting all dependents of the five libraries, we remove the duplicates as some of the dependent projects use more than one library (e.g., pandas and NumPy). % I want include this clarification, because if we add the number of dependents it is closer 2.8 mill. Then, a reviewer might wonder how we came to 800K from 2.8 mill. The way I wrote it previous is confusing.
After removing duplicates, we find 869,544  projects that use at least one of the five libraries. 
We collect the metadata of all these dependent repositories; specifically, stargazer count, number of contributors, programming language, visibility (public or private), archived or not, fork or not, project creation date, and the last committed date. 
From the collected metadata, we remove projects whose main programming language is not Python. 
We filter out projects that are either private or archived. 
We also filter out forks. 
After removing repositories using the above filters, we sort the remaining repositories in descending order based on stargazer count, number of contributors, and the project’s age. We calculate the age by subtracting the created date from the last committed date. We then select the top 100 projects from this ranked list to analyze their commits. 
From the selected 100 repositories, we select candidate commits that may contain API misuses for manual evaluation.
Specifically, a commit that contains changes to an API usage of one the five libraries could potentially be a fix for a misuse. 

%\input{tables/tab-gh-project-count.tex}

% \sn{this paragraph is a candidate for shortening if we need space.}
% To find these candidate commits, we collect a list of all APIs from each library through Python's inspect module \cite{inspect}.
% To find API changes in commits, we use PyDriller~\cite{pydriller} to mine commits in the selected projects. PyDriller provides file names that have been modified including the modified source lines with their line numbers. 
% We first ignore commits where the number of modified lines in a file is more than ten as previous studies observed that fixing API misuses involved small number of edits\cite{Detector-Eval-Amann:2019,dl-api-misuse-llm:2024}.
% For the remaining commits, we then parse the changed Python files (the version before the change and after the change) to find usages of the libraries we are interested in. 
% Our parser analyzes the abstract syntax tree (AST) and collects information such as import statements and function calls.
% The parser returns any used APIs from our target libraries, along with their line numbers. If any of these line numbers is a modified line according to PyDriller, we consider this commit as a candidate for manual analysis. 
% To reduce the number of returned candidate commits, we filter out commits that are likely fixing a typo than changing the API usage. Specifically, we use regular expressions to determine if the commit message is one liner, and the message contains ``fix'' or ``correct'' with ``typo''. 

To find candidate commits, we use PyDriller~\cite{pydriller} to first detect the modified Python files in each commit along with the modified line numbers. 
We ignore changed files with less than 10 changed lines, as previous studies observed that fixing API misuses typically involved small number of edits~\cite{Detector-Eval-Amann:2019,dl-api-misuse-llm:2024}. 
We also filter out commits that are likely fixing a typo than changing the API usage by using a regular expression to determine if the commit message is a one-liner, and the message contains ``fix'' or ``correct'' with ``typo''.
For the remaining commits, we parse the changed Python files to find if any of the modified lines contain API usages of the target libraries. 
%Our parser employs an AST-based approach, examining each library's list of APIs to extract relevant API occurrences along with their corresponding line numbers. 
%If any of these line numbers coincide with modified lines returned by PyDriller, we consider this commit as a candidate for manual analysis. 

Overall, we find 3,447 candidate commits originating from 76 of the analyzed projects. 
%24 projects did not have any candidate commits. 
Table \ref{tab:gh_commits} shows the number of candidate commits per library. In these commits, we find that APIs such as \texttt{numpy.array} or \texttt{pandas.DataFrame} were modified over hundred times while APIs such as \texttt{seaborn.heatmap} or \texttt{matpl\-otlib.Axes3D} were modified less than ten times. Additionally, many APIs were modified only once. To prioritize the discovery of unique API misuses, we focus on selecting commits that modify different APIs. For APIs that appear in less than three candidate commits, we include all related commits for those changes. However, if an API appears in more than three candidate commits, we randomly select three commits from different projects.  
From the 3,447 candidate commits, we first select commits that modify unique APIs, resulting in identifying 1,295 diverse commits (Table \ref{tab:gh_commits} column 3). These commits contain 574 unique APIs that were modified. We manually analyze \totalghcommits randomly selected diverse commits (Table \ref{tab:gh_commits} column 4) to identify API misuses, satisfying a 95\% confidence interval and 5\% error margin. 

\begin{table}[t!]
	\centering
	\caption{Number of candidate commits for each library.\vspace{-0.3cm}}
	\resizebox{0.45\textwidth}{!}{
	\begin{tabular}{lrrrrr}
		\toprule
		\multicolumn{1}{l}{\textbf{Library}} & 
		\multicolumn{1}{c}{\textbf{\# of all}} &
		\multicolumn{1}{c}{\textbf{\# of diverse}} &
		\multicolumn{1}{c}{\textbf{\# of reviewed}} &
		\multicolumn{1}{c}{\textbf{\# of unique}} &
		\multicolumn{1}{c}{\textbf{\# of reviewed}}\\

		&
		\multicolumn{1}{c}{\textbf{commits}} &
		\multicolumn{1}{c}{\textbf{commits}} &
		\multicolumn{1}{c}{\textbf{commits}} &
		\multicolumn{1}{c}{\textbf{APIs}} &
		\multicolumn{1}{c}{\textbf{APIs}}\\
		\midrule
		pandas 			& 390		& 185 	& 50 (27\%)		& 92	& 37 (40\%)	\\
		NumPy 			& 2273 		& 431  	& 103 (24\%)	& 184	& 74 (40\%)	\\
		scikit-learn 	& 545 		& 528  	& 143 (11\%)	& 210	& 84 (40\%)	\\
		Matplotlib 		& 234 		& 146  	& 57 (39\%)		& 80	& 42 (53\%)	\\
		seaborn 		& 5 		& 5    	& 5  (100\%)	& 8		& 8	 (100\%)	\\
		\midrule
        Total 			& 3447		& 1,295 	& 358 (28\%)	& 574	& 245 (43\%)	\\
		\bottomrule	
		
	\end{tabular}
	}
	\label{tab:gh_commits}
	\vspace{-0.5cm}
\end{table}

% \todo{depends on next annotation rounds this section may get longer}

\vspace{-0.3cm}
\subsection{Data Analysis}

We rely on manually analyzing a given candidate \so{} post or GitHub commit to determine if it contains an API misuse.
We create a coding guideline to streamline this process.
In the subsections below, we first describe how we iteratively developed our coding guideline using \so{} data and then describe the developed criteria.
Finally, we discuss our closed coding process for labeling the identified misuses.
% Table \ref{tab:annotation_ex} shows examples of annotating \so{} questions and GitHub commits.

\vspace{-0.2cm}
\subsubsection{Developing a coding guideline:} 
Stack Overflow threads typically have a wealth of information, including problem descriptions, code examples, stack traces, comments, and answers. 
Thus, we use the process of analyzing \so{} posts to iteratively develop and refine our coding guide for confirming API misuse.

Specifically, we first conduct the manual annotation process of \so{} posts iteratively using three authors of this paper. We start with an initial coding guide on how to identify API misuses. After each iteration, annotators discuss any disagreements that occur in that round. If a resolution surfaces any new information that has not been included in the initial coding guide, we improve the coding guideline to be more explicit. In each round of the first five rounds, we select six \so{} questions for each library, totaling 30 questions per round.  
After each round, we calculate the agreement score for determining a misuse  using Fleiss kappa score~\cite{Kappa:2015}. 
In the first three rounds, we reach moderate agreement ($>$0.5)~\cite{Kappa:1977}. In round four, our agreement was fair (0.37)~\cite{Kappa:1977}.
After revising the coding guide, in the fifth round, we received an almost perfect agreement score (0.85)~\cite{Kappa:1977} which indicates that the annotators achieved a solid understanding on how to identify and distinguish API misuses and that the instructions in the code guide are stable. 
After the fifth round, the same three authors continue analyzing the remaining threads but with only two annotators per thread. Overall, we maintain a moderate agreement ($>$0.5), measured through a pair-wise kappa score~\cite{Kappa:1977}, during the last round of annotation. 
Each pair of annotators resolved their disagreements, sometimes involving the third annotator if needed.
Given the stability of the coding guideline at this point, we proceed to assign only one annotator per selected GitHub commit. 
We then perform an additional verification step where we create a minimal reproducible example for each identified misuse to confirm that our understanding of the misuse is correct.
%% \ag{I think we can remove this table as our description is clear.}
% \input{tables/tab-annotation-ex.tex}

\vspace{-0.2cm}
\subsubsection{Coding guideline \& coding process}
Our exact coding guideline is part of our artifact. 
When analyzing a \so{} thread, we make use of the question title, description, provided code snippets, as well as the answers in the thread.
We discard \so{} posts that mainly seek help with implementing a particular functionality (e.g., \textit{best way to plot two graphs on the same axis}) or understanding code/concepts (e.g., \textit{why does a specific function use a particular approximation method}). 
On the other hand, when analyzing GitHub commits, we make use of the commit message to understand the purpose of the change, as well as reading through the modified code lines.
In some commits where a bug issue or pull request is linked in the commit message, we also read through the conversation there to get more context about the fix.
We now describe some of the important criteria for determining a misuse.

If the problem is related to the user's intention (e.g., they want the legend on the right not the left), we do not consider this as a misuse (Definition \ref{def:misuse}). Additionally, if the fix is related to managing API deprecation, we do not consider this as a misuse (Definition~\ref{def:usageconstraint}).
To differentiate API misuses from normal Python (type) errors, we do not consider problems that are caught by the Python type system as a misuse (e.g., passing float instead of integer). 
%For example, a developer passing a float value as an argument instead of an integer is a typing problem that can be handled by Python, and is thus not an API misuse. 

Once we determine that a thread/commit contains a misuse, we determine the root cause of the problem and record its description as free text characteristics.
We use these noted characteristics to later categorize our collected misuses using Wei \etal~\cite{dl-api-misuse-llm:2024}'s taxonomy. 
%We adopt Wei \etal's taxonomy \cite{dl-api-misuse-llm:2024} to illustrate that the characteristics of API misuses in data-centric libraries extend beyond deep learning libraries like PyTorch or TensorFlow.
% We also record the symptom that was reported to be observed (Table \ref{tab:annotation_ex} column 8).
% We also attempt to identify and record the API usage constraint that was violated and resulted in the misuse.
% We refer to the documentation to verify the presence of the explained or quoted \cav.
% If the \cav{} is documented, we record the directive along with a link to the source (Table \ref{tab:annotation_ex}, column 7).

%% \ag{Without annotation table}
We also record the symptom that was reported to be observed.
Finally, we refer to the library documentation to identify whether the API usage constraint that was violated is documented in the API.
If the \cav{} is documented, we record the directive along with a link to the source.

% In some cases, \so{} answers/comments include an explanation quoting text directly from API documentation or provide a link to the documentation \cite{API-KG-Li:2018}; in other cases, they provide their own guidelines without quoting documentation. In both cases, we refer to the documentation to verify the presence of the explained or quoted \cav.
% If the \cav{} is documented, we record the directive along with a link to the source (Table \ref{tab:annotation_ex}, column 6). While some \cavs{} are documented, some are implicit (see Definition~\ref{def:directive}). In that case, we provide a free-text usage guideline that summarizes the problem according to the thread; we express this guideline as an \cav{} sentence that \textit{could potentially} appear in the documentation (but in this case does not). 
% Since GitHub commits normally do not have conversations like \so{} comments, we refer to the API documentation to find explicit \cav{}, or we provide a guideline in the absence of documented \cav. 
  
\vspace{-0.2cm}
\subsubsection{Categorizing API misuses}
We use closed coding \cite{charmaz2014constructing} to label the confirmed misuses according to  Wei \etal~\cite{dl-api-misuse-llm:2024}'s taxonomy described in Section~\ref{sec:background-taxonomy}.
Specifically, we use our recorded notes to determine each misuse's type (program element and violation), root cause, and symptom.
We allow extensions to the taxonomy if we find any misuses that do not fit into one of the existing categories.

\vspace{-0.3cm}
\section{Characteristics of API Misuses in Data-centric Libraries}
We find that approximately 8\% of the \so{} posts and 5\% of the GitHub commits we analyzed contain API misuses.
In total, we identify \finalmisuses misuses, with \somisuses identified from Stack Overflow and \ghmisuses from GitHub. Our data set contains misuses of \distinctAPIs unique APIs. Table \ref{tab:misuse-lib-stats} provides a summary of the number of misuses, and their distribution across the five libraries.
We now describe the details of the misuses we found, using the taxonomy in Figure~\ref{fig:taxonomy}. 

%Among the misuses identified in Stack Overflow posts, over half were related to visualization libraries, with scikit-learn accounting for a quarter of the total, and the remaining misuses attributed to pandas and NumPy. Conversely, in GitHub, no misuses were found for seaborn, while Matplotlib accounted for 20\% of the misuses. The remainder of misuses are equally distributed between pandas, NumPy, and scikit-learn.

% \sn{idea: "Shape mismatch is a unique type of API misuse in deep learning libraries because these libraries heavily rely on tensor computations." --> but we observed some form of shape mismatch too? like incorrect dataframe dimensions, column types etc. So that's a way to frame our findings that it's not necessaruly only a DL thing but beyond that} 

\begin{table}[t!]
	\centering
	\caption{Statistics of identified misuses.\vspace{-0.3cm}}
	\resizebox{0.25\textwidth}{!}{
	\begin{tabular}{lrrr}
		\toprule
		\multirow{2}{*}{\textbf{Library}} &
		\multicolumn{2}{c}{\textbf{Data Source}} & 
		\multirow{2}{*}{\textbf{Total}}\\
		\cmidrule{2-3}
		&
		\multicolumn{1}{c}{\textbf{SO}} &
        \multicolumn{1}{c}{\textbf{GH}}&
        \\
		\midrule
		pandas          &  4  		&  1  	& 5 (10\%) \\
		NumPy           &  2 		&  11  	& 13 (27\%) \\
		scikit-learn    &  7 		&  6  	& 13 (27\%) \\
		Matplotlib      &  7 		&  3  	& 10 (20\%) \\
		seaborn         &  8 		&  0 	& 8 (16\%) \\
		\midrule
        Total           & 28      	& 21    & 49 (100\%)  \\
		\bottomrule	
		
	\end{tabular}
	}
	\label{tab:misuse-lib-stats}
	\vspace{-0.5cm}
\end{table}

\vspace{-0.3cm}
\subsection{API misuse types}
\label{res:types}

The red boxes in Figure~\ref{fig:taxonomy} show the additions we made to the taxonomy.
Specifically, we find misuses that occur due to violating the expected order of execution, which we add.
We also observe that some misuses are identified by problems in the data being processed. For example, the misuse in Figure \ref{fig:seaborn-example} is \emph{data dependent}, because if the column `color' passed to \texttt{hue} has categorical data, the API works correctly. Therefore, we extend the taxonomy to include data dependency as a dimension for describing misuse type.
% (Table \ref{tab:annotation_ex}, column 9 shows the value for the provided examples).

We now discuss the types of API misuses we observe based on this updated taxonomy.
Table \ref{tab:type_distribution} shows the distribution of misuse types in our data set compared to the deep learning misuse statistics~\cite{dl-api-misuse-llm:2024}, which are shaded gray.
Since we do not consider deprecations as misuses in our study, the \textit{outdated} column in Table \ref{tab:type_distribution} is empty for our data. 
The table also shows how many of the misuses are data dependent, for each misuse type.
Overall, out of the \finalmisuses misuses, \datadep (\datadepPerc) are data dependent. 

Wei et al.~\cite{dl-api-misuse-llm:2024} found that 51\% of the deep learning misuses involved API methods, 35\% involved API parameters, while 13\% involved API conditions.
In our study, \methodPerc of the misuses involve API methods, \paramPerc involve API parameters, while \conditionPerc involve API conditions. 
The higher percentage of API parameter misuses in our study can be attributed to the diverse range of data types accepted by the libraries we analyzed, coupled with the specific constraints imposed on the data within a given context. For example, Matplotlib's \texttt{text} method accepts any data type for the \texttt{x} and \texttt{y} parameters, but the data type must match the previously set axes type. 
% The grayed rows indicate the types of misuses that Wei \etal~\cite{dl-api-misuse-llm:2024} found in their study of deep learning libraries. 
We now discuss the overall API misuse types we observe in terms of combination of program element and violation type.

\begin{table}[t!]
    \centering
    \caption{API misuse types of the data-centric libraries we study compared to deep learning libraries~\cite{dl-api-misuse-llm:2024}, shown in gray.\vspace{-0.3cm}}
    \resizebox{0.47\textwidth}{!}{
        \begin{tabular}{lllrrrrrr}
            \toprule

            &&&
            \multicolumn{1}{c}{Missing} &
            \multicolumn{1}{c}{Redundant} &
            \multicolumn{1}{c}{Replacement} &
            \multicolumn{1}{c}{Ordering} &
            \multicolumn{1}{c}{Outdated} &
            \multicolumn{1}{c}{\textbf{Total}} \\

            \midrule

            \multirow{3}{*}{\textbf{API method}} & \multirow{2}{*}{\textbf{This study}}    & \textbf{Total}    & 9 (18\%) & 6 (12\%) & 4(8\%)    & 2 (4\%) & - & 21 (43\%) \\
                                                                                             \cmidrule{3-9}
                                                 &                                           & \textbf{DD}       & 4 (15\%) & 5 (19\%) & 1 (4\%)   & 1 (4\%) & - & 11 (42\%) \\
                                                                                             \cmidrule{2-9}
                                                 & \multicolumn{2}{l}{\textbf{DL only~\cite{dl-api-misuse-llm:2024}}} & \cellcolor{gray!25} 113 (13\%) & \cellcolor{gray!25} 138 (15\%) & \cellcolor{gray!25} 130 (15\%) & \cellcolor{gray!25} - & \cellcolor{gray!25} 74 (8\%) & \cellcolor{gray!25} 455 (51\%) \\
            \midrule
            \multirow{3}{*}{\textbf{API parameter}} & \multirow{2}{*}{\textbf{This study}} & \textbf{Total}    & 9 (18\%) & 1(2\%)  & 15 (31\%) & -      & - & 25 (51\%) \\
                                                                                             \cmidrule{3-9}
                                                    &                                          & \textbf{DD}     & 5 (19\%) & -       & 10 (37\%) & -      & - & 15 (56\%) \\
                                                                                             \cmidrule{2-9}
                                                    & \multicolumn{2}{l}{\textbf{DL only~\cite{dl-api-misuse-llm:2024}}}& \cellcolor{gray!25} 88 (10\%) & \cellcolor{gray!25} 56 (6\%) & \cellcolor{gray!25} 60 (7\%) & \cellcolor{gray!25} - & \cellcolor{gray!25} 115 (13\%) & \cellcolor{gray!25} 319 (36\%) \\
            \midrule
            \multirow{3}{*}{\textbf{API condition}} & \multirow{2}{*}{\textbf{This study}} & \textbf{Total}    & 3 (6\%)  & -       & -         & -      & - & 3 (6\%) \\
                                                                                             \cmidrule{3-9}
                                                    &                                          & \textbf{DD}     & 1 (4\%)  & -       & -         & -      & - & 1 (4\%) \\
                                                                                             \cmidrule{2-9}
                                                    & \multicolumn{2}{l}{\textbf{DL only~\cite{dl-api-misuse-llm:2024}}}& \cellcolor{gray!25} 43 (5\%) & \cellcolor{gray!25} 29 (3\%) & \cellcolor{gray!25} 17 (2\%) & \cellcolor{gray!25} - & \cellcolor{gray!25} 28 (3\%) & \cellcolor{gray!25} 117 (13\%) \\

            \bottomrule
            \multicolumn{9}{l}{DD=data dependent, DL=deep learning}\\
        \end{tabular}
    }
    \vspace{-0.5cm}
	\label{tab:type_distribution}
\end{table}

\noindent\textit{\textbf{Missing API Method:}}
In general, missing an API call is one of the most commonly discussed type of API misuse in the literature~\cite{FUM:2022, misuse-wild-Li:2021, Detector-Eval-Amann:2019, dynamic-misuse-He:2023}.
We observe that \checknum{nine} (\checknum{18\%}) misuses of our studied libraries result from developers missing necessary calls to an API.

When examining misuses in deep learning applications, Wei \etal~\cite{dl-api-misuse-llm:2024} found that failing to call APIs such as \texttt{flatten} causes shape mismatch errors. While shape mismatch errors are unique type of API misuse in deep learning libraries, as they are heavily reliant on tensor computations, we also observe similar misuses in our data set. For example, in pandas, failing to call \texttt{pivot} on a pandas dataframe before passing it to \texttt{heatmap} results in a runtime error, because the input is not in wide format as \texttt{heatmap} expects.
It is important to note that this misuse is data dependent and only occurs when the data is in long format.
% We only observe one other data dependent missing API method misuse. 

% We find that \checknum{9\%} of these missing API calls are data dependent.
% For example, seaborn's \texttt{heatmap} method expects the input data in wide-format. When data is in long-format (i.e. like a pandas dataframe), developers need to first call \texttt{pivot} to change it to the correct format.
% Thus, missing a \texttt{pivot} call is a misuse \textit{only} if the data is in long format.

% \paragraph{Redundant API Method:}
\noindent\textit{\textbf{Redundant API Method:}}
An API call is redundant if it is an API that the developer should not call in the given context and calling it would lead to an unexpected outcome or has no effect. 
%Note that previous API misuse work referred to redundant API calls as \textit{superfluous}~\cite{Detector-Eval-Amann:2019}.
Figure \ref{fig:lmplot} shows an example of a redundant call to \texttt{FacetGrid}. Since \texttt{lmplot} internally uses \texttt{FacetGrid}, users could simply use \texttt{lmplot} (Line 6) without passing it to \texttt{FacetGrid} (Line 4, 5). 
In this case, the extra call is not harmless, using the combination shown in Figure~\ref{fig:lmplot} Line 4, 5 actually causes a run-time error. 
Notably, out of the \checknum{six} misuses in this category, \checknum{five} are data dependent. 
% Among all data-dependent misuses, the proportion of redundant API method is the third highest category, accounting for \checknum{22\%} of total data-dependent misuses.

\begin{figure}[t]
\begin{lstlisting}[style=pythoncode]
import seaborn as sns

tips = sns.load_dataset('tips')
- g = sns.FacetGrid(data=tips, col='time', row='sex')
- g.map(sns.lmplot, 'total_bill', 'tip')
+ sns.lmplot(data=tips, x='total_bill', y='tip', col='time', row='sex')
\end{lstlisting}
	\vspace{-0.3cm}
	\caption{Redundant API call to seaborn's \texttt{FacetGrid} when using \texttt{lmplot}}
	\label{fig:lmplot}
	\vspace{-0.6cm}
\end{figure}

% \paragraph{API Method Replacement:}
\noindent\textit{\textbf{API Method Replacement:}}
We find \checknum{four} misuses (\checknum{8\%}) where the developer used an incorrect API; only \checknum{one} of them are data dependent. For example, when using scikit-learn's \texttt{StandardScaler}, calling \texttt{transform} before it trains on a sample of data is incorrect. The correct method to call is \texttt{fit\_transform} \cite{sk-fit-transform}. %None of these four misuses are data dependent.

% \paragraph{API Method Ordering:}
\noindent\textit{\textbf{API Method Ordering:}}
Some libraries expect APIs to be called in a specific order. For example, seaborn expects developers to set the axis tick labels \textit{before} applying any formatting to the labels. Otherwise, the formatting is simply not applied. 
We find \checknum{two} misuses (\checknum{4\%}) that belong to this category, with only \checknum{one} of them being data dependent.
While Wei \etal~\cite{dl-api-misuse-llm:2024} did not observe violation type related to call order, the FUM taxonomy~\cite{FUM:2022} uses the label ``Method call sequence'' to signify misuses in this category.

\noindent\textit{\textbf{Missing API Parameter:}}
The second most common type of misuse is missing API parameter, representing \checknum{18\%} of our misuses.
APIs can specify the parameters they accept, which can be required or optional, as well as keyword based or position based. 
When a developer does not pass a required parameter, Python would complain and issue a run-time error. 
Recall that we do not consider such cases since they are general Python programming errors that the interpreter can easily detect.
However, there are cases where other code context implies the necessity of setting a particular parameter (or its value). This is what this API misuse type refers to.

When creating tensors in TensorFlow, the parameter \texttt{dtype} is optional. If a subsequent API requires a specific data type for the input tensor, failing to set \texttt{dtype} appropriately can lead to unexpected results and propagate errors throughout the program \cite{dl-api-misuse-llm:2024}.
We observe a similar case in pandas where \texttt{replace(...)}~\cite{pd-replace} accepts either True or False for the parameter \texttt{inplace}. 
The default value of \texttt{inplace} is False, which means that \texttt{replace} would return a new data frame with the replaced values.
Accordingly, a misuse occurs if the code does not have an assignment operation like \texttt{new\_df = df.replace(...)} and the developer continues to use the old \texttt{df}, incorrectly assuming it has the replaced values.
% In this case, the developer missed setting the value of the \texttt{inplace} parameter to True, resulting in an API misuse.
We also observe a case where the missing API parameter depends on another related parameter.
For example, in Figure \ref{fig:param_corr}, the developer tries to set \texttt{norm\_hist=False} (Line 4) without setting \texttt{kde=False}. Line 5 shows the correct setting of these related parameters. Without that, the y-axis show density instead of counts. We find that \checknum{19\%} of the data-dependent misuses are missing API parameter misuses, making it the second highest category in misuses that are data dependent.

\begin{figure}[t]
\begin{lstlisting}[style=pythoncode]
import seaborn as sns 

data = sns.load_dataset('tips')
- sns.distplot(data.tip, norm_hist=False)
+ sns.distplot(data.tip, norm_hist=False, kde=False)
\end{lstlisting}
	\vspace{-0.3cm}
	\caption{Missing API parameter for seaborn's \texttt{distplot}.}
	\label{fig:param_corr}
	\vspace{-1cm}
\end{figure}

\noindent\textit{\textbf{API Parameter Replacement:}}
We find that API parameter replacement is the most common type of misuse in our data (\checknum{31\%}).
This occurs when a method-call parameter is used incorrectly where a different value should be passed.
Examples of incorrect usage includes passing incorrect data formats or unsupported values.%, or incorrect types. Note that, we do not consider simple type mismatches that Python's type system will catch (Definition \ref{def:misuse}). 

For instance, in Matplotlib's \texttt{datestr2num(...)} \cite{datestr2num_mpl} method, the expected input date format is month-first (e.g., 04-21-2021). Providing a day-first format results in incorrect output due to improper conversion. Another example is when invoking \texttt{set\_style(...)} \cite{sb_setstyle} from seaborn to modify the aesthetics of a graph, developers can specify a set of key-value pairs (as a Python dictionary) from a predefined list of supported pairs. If an unsupported value is passed, seaborn silently ignores it, resulting in no change to the graph's appearance. As a last example, according to Matplotlib's documentation, calling \texttt{pyplot.text(x, y, s)} \cite{mpl_text} to add a text label to a graph requires that the values of parameters \texttt{x} and \texttt{y} match the types of the corresponding axes (which would be determined from previous API calls while setting up the graph). A run-time error occurs when the argument type does not align with the axes type.
Wei \etal~\cite{dl-api-misuse-llm:2024} also observed API parameter replacement in deep learning APIs, where developers explicitly set the \texttt{dtype} parameter to \texttt{float32} to reduce computational costs.

We note that only 7\% of the deep learning API misuses are API parameter replacements~\cite{dl-api-misuse-llm:2024}, versus \checknum{31\%} in our data. 
We believe that the reason for this difference is that deep learning libraries have a more restricted set of data types and values that can be passed to their APIs, while the libraries we studied (e.g., Matplotlib and seaborn) have a wider range of data types and values that can be passed to their APIs.
We also note that \checknum{37\%} of API misuses that are data dependent belong to this misuse category.
% \paragraph{Redundant API parameter:}
% Misuses in this category occur when a developer unnecessarily sets a parameter. For example, passing a string to \texttt{label} parameter of \texttt{matplotlib.pyplot.colorbar} while calling \texttt{matplotlib.pyplot.title} is redundant, because it causes an unexpected behavior when running the continuous integration system\footnote{https://github.com/scikit-learn/scikit-learn/commit/eaf97bf}.
% \sn{not really clear writing wise. Additionally, is it that it has to result in incorrect behavior (isn't that the consequence?) or is it just that it is unnecessary?}
% \ag{Went back to analyze this again. Don't think this is a misuse as the plt.title and plt.colorbar(label=...) does not interfere with each other. For some reason, the CI system of this project failed. That does not necessarily mean this is a misuse.}

\begin{findingenv}{Summary}{finding:misuse-summary}
API parameter replacement is the most common misuse type in our data, followed by missing API parameter. 
Misuses in these two categories account for \checknum{49\%} of the total misuses. 
Overall, we observe all API misuse types found in deep learning libraries, and also discover the new dimension of data-dependent misuses.
\end{findingenv}

%%%%%%%%%%%%%%%%%%%%%%%%%%%%%%%%%%%%%%%%%%%%%%%%%%%%%%%%%%%%%%%%%%%%%%%%%%%%%%%%%%%%%%%%%%%%%%%%%%%%%%%%%%%%%%%%%%%%
\vspace{-0.3cm}
\subsection{API Misuse Root Causes}
We find that we could describe most of the root causes of our misuses using the existing taxonomy.
However, we also observe the need for a new category of root causes, namely \textit{configuration errors}, to allow us to capture 12 of our observed misuses. 
Table \ref{tab:root-causes} shows the distribution of root causes of data-centric API misuses compared to deep learning API misuses.

% \paragraph{Data conversion errors:}
\noindent\textit{\textbf{Data conversion errors:}}
Data conversion errors are the number one root cause of misuses of our data set (\checknum{39\%}), while they were the second highest for deep learning misuses since they were preceded by device management errors.
As an example, when utilizing visualization APIs such as seaborn's \texttt{lineplot}, providing column vectors for both the \texttt{x} and \texttt{y} parameters leads to a runtime error, as the API expects a 1D array. In another case, scikit-learn's \texttt{OneHotEncoder} requires the input dataframe column to be of uniform data type; otherwise, it cannot internally convert the data.

% \paragraph{Null reference errors:}
\noindent\textit{\textbf{Null reference errors:}}
Similar to findings of Wei \etal~\cite{dl-api-misuse-llm:2024} (4\%), we also observe a small percentage (\checknum{10\%}) of null reference errors. 
% Notably, some of these misuses present interesting scenarios that may be easily identifiable to developers. For example, in scikit-learn's \texttt{ColumnTransformer}, transformations do not occur in the order of they are specified in the input array. Therefore, as shown in Figure \ref{fig:col_trans}, if the \texttt{OneHotEncoder} (Line 20) is applied before the creationg of \texttt{healthy} column (Line 18), a null reference error may result when the \texttt{ColumnTransformer} tries to apply the transformation on the \texttt{healthy} column.

% \begin{figure}[t]
% 	\lstinputlisting[style=pythoncode]{listings/col_tranf_diff.py}
% 	\vspace{-0.3cm}
% 	\caption{Null reference error in scikit-learn's \texttt{ColumnTransformer}.}
% 	\vspace{-0.4cm}
% 	\label{fig:col_trans}
% \end{figure}

% \ag{This occupies a lot of space. We may choose to remove the example as the category itself has fewer misuses. However, I don't think we can explain this example without the listing.}

% \paragraph{Algorithm errors:}
\noindent\textit{\textbf{Algorithm errors:}}
We observe a higher proportion (\checknum{20\%}) of misuses whose root cause is algorithm errors.
In the deep learning data set, the majority of misuses in this category were attributed to division by zero, often due to developers neglecting to pass a parameter (a small floating-point value) to mitigate such numerical errors \cite{dl-api-misuse-llm:2024}. In our data set, the algorithm problems varied. For example, scikit-learn's \texttt{roc\_curve} API being incorrectly applied to multi-class classification, resulting in inaccurate results, even though the API is designed for binary classification.

% \paragraph{Device management errors:}
\noindent\textit{\textbf{Device management errors:}}
In deep learning libraries, device management errors are the highest root cause, with the vast majority being specific to CPU vs. GPU usage. 
We observe only \checknum{one} device management error where a developer forgot to call \texttt{pyplot.close()} when plotting figures in a loop. 
We, expectedly, do not observe CPU/GPU usage related device management errors.
%While this is not a typical device management error that observed in deep learning APIs, Wei \etal~\cite{dl-api-misuse-llm:2024} defined this category as misuses related to hardware and resource management, which includes memory issues like this.

% \paragraph{Configuration errors:}
\noindent\textit{\textbf{Configuration errors:}}
There were many misuses whose root cause did not fit within the existing taxonomy.
Accordingly, we derive this new category that refers to misuses resulting from incorrect or missing internally expected configuration settings within an API. %, leading to unexpected or incorrect behavior, excluding errors related to device management. 
%%example from SO thread 59415689
For example, seaborn's \texttt{countplot} does not set the saturation values to the provided input values unless the \texttt{saturation} argument is set to 1; otherwise, an incorrect internal library/API configuration setting would be used.
%\sn{I am not entirely convinced that fig 1 is a configuration error or at least the best example to use here so tried to find another simple example}
%For example, the misuse shown in Figure \ref{fig:seaborn-example} does not fit into any of the existing root causes. Given the input data, calling \texttt{set\_pallette} is an incorrect API configuration. 
We find that approximately \checknum{24\%} of the misuses, mostly visualization API misuses, are due to such internally expected configurations.

% \paragraph{Other:}
\noindent\textit{\textbf{Other:}}
We find \checknum{two} misuses (\checknum{4\%}) that did not clearly fall into any of the root cause categories. One is the misuse in Figure \ref{fig:lmplot}, where calling \texttt{FacetGrid} is redundant and problematic, because \texttt{lmplot} internally calls \texttt{FacetGrid}.

\vspace{-0.2cm}
\begin{findingenv}{Summary}{finding:root-cause-summary}
We find that data conversion errors in the additional \dsml{} libraries we study are even more prominent than deep learning libraries, while device management errors are naturally less.
%	We observe that data-centric API misuses share common root causes with deep learning API misuses, such as data conversion errors, while device management errors appear to be dominant in deep learning libraries. 
	We also find a new category of root causes, \emph{configuration errors}, encompassing \checknum{$~$24\%} of the observed misuses.
\end{findingenv}

%%%%%%%%%%%%%%%%%%%%%%%%%%%%%%%%%%%%%%%%%%%%%%%%%%%%%%%%%%%%%%%%%%%%%%%%%%%%%%%%%%%%%%%%%%%%%%%%%%%%%%%%%%%%%%%%%%%%
\vspace{-0.3cm}
\subsection{API Misuse Symptoms}

% \paragraph{Program crash:}
\textit{\textbf{Program crashes:}}
% Similar to Wei \etal~\cite{dl-api-misuse-llm:2024}, we also find that \misuseWithErrorsPerc of our observed misuses cause program crashes. 
Similar to Wei \etal{} (36\%)~\cite{dl-api-misuse-llm:2024}, we also find that program crashes is the most frequent misuse symptom (\misuseWithErrorsPerc).

\begin{table}[t!]
	\centering
	\caption{Distribution of root causes of data-centric API misuses and deep learning API misuses \cite{dl-api-misuse-llm:2024}.\vspace{-0.3cm}}
	\resizebox{0.4\textwidth}{!}{
	\begin{tabular}{lrrr}
		\toprule
		\multicolumn{1}{l}{\textbf{Category}} &
		\multicolumn{1}{c}{\textbf{\# (\%) data-centric}} & 
        \multicolumn{1}{c}{\textbf{\# (\%) deep learning}~\cite{dl-api-misuse-llm:2024}} \\
		
		\midrule
		Data conversion error       &  19 (39\%)    &  246 (28\%)     \\
		Algorithm error             &  10 (20\%)     &  88 (10\%)     \\
		Null reference error        &  5 (10\%)      &  33 (4\%)     \\
		Device management error     &  1 (2\%)      &  337 (38\%)     \\
		Configuration error         &  12 (24\%)     &  -              \\
        Other errors                &  2 (4\%)      &  10 (1\%)      \\
		\bottomrule	
		
	\end{tabular}
	}
	\vspace{-0.5cm}
	\label{tab:root-causes}
\end{table}

% \vspace{-0.3cm}
% \paragraph{Unexpected output:}
\noindent\textit{\textbf{Unexpected output:}}
Despite our smaller dataset, we observe a higher proportion of misuses that result in unexpected output (\misuseWithIncorrectOutputPerc) compared to the 24\% of API misuses found by Wei \etal{} \cite{dl-api-misuse-llm:2024}. Specifically, we find that \misuseWithIncorrectOutput misuses did not result in any runtime errors but instead silently produced incorrect or unexpected outputs which could propagate through the program without a notice.
% Wei \etal's~\cite{dl-api-misuse-llm:2024} found that only approximately 24\% of the API misuses of deep learning libraries resulted in unexpected output. 

Out of the \misuseWithIncorrectOutput misuses with unexpected output that we observe, \checknum{11} are related to APIs from the two visualization libraries while the remaining \checknum{6} instances are from scikit-learn and pandas. Figure \ref{fig:seaborn-incorrectgraph} is an example of incorrect output resulting from misusing visualization library APIs. 
We also observe cases where misusing the visualization library's APIs resulted in incorrect axis tick labels, incorrect axis scales (count vs. normalized values), and incorrect plot styles such as not displaying an axis grid. The incorrect outputs caused by misusing APIs from non-visualization libraries varied. For example, when calling \texttt{replace()} on pandas dataframe, the API returns a dataframe with replaced values. Failing to assign the return dataframe produces incorrect output (that is harder to spot than an incorrect visualization for example), because the developers continue to process the dataframe with initial values. 

% \vspace{-0.3cm}
% \paragraph{Low Efficiency:}
\noindent\textit{\textbf{Low efficiency:}}
Only \misuseWithMemoryPerc of the misuses we observed result in low efficiency issues related to performance or memory usage. In contrast, Wei et al.'s findings \cite{dl-api-misuse-llm:2024} indicate that 32\% of misuses in deep learning APIs lead to poor performance. We note that in their data, device management errors accounted for the majority of misuse root causes (38\%), which explains why they find a higher proportion of misuses that result in low efficiency since 
configuring the correct device (i.e., CPU vs GPU) could result in faster training, considering the high computation cost of deep learning models.
In our studied libraries, the low efficiency symptoms are related to the misuse of APIs that are computationally expensive, when specific configurations are not set. For example, when calling \texttt{intersect1d} in NumPy, the parameter \texttt{assume\_unique} should bet set to \texttt{True} if the input arrays are unique, speeding up calculation.

% \vspace{-0.3cm}
% \paragraph{Return warning:}
\noindent\textit{\textbf{Return warning:}}
Only two misuses in our dataset
%—one from pandas and one from Matplotlib—
resulted in warnings. One pandas misuse is due to chain indexing, while one Matplotlib misuse is caused by calling the \texttt{legend} function when there was no legend to display. 
We note that Wei \etal~\cite{dl-api-misuse-llm:2024} observe much more warnings, because most of their observed warnings were deprecation warnings, which we do not consider.% that the higher number (64) of misuses in Wei \etal~\cite{dl-api-misuse-llm:2024}'s study  is attributed to their inclusion of deprecation as a misuse. Since we did not consider deprecation in our study, misuses that result in warnings appear less frequently in our data.

% It is worth noting that Wei et al. \cite{dl-api-misuse-llm:2024} observed a much higher number of misuses (64) that returned warnings, although considerably less than other impact categories. This is partly due to their inclusion of deprecations as misuses, as warnings are generated by the Python interpreter when an API is deprecated. Since we did not consider deprecation in our study, misuses that result in warnings appear less in our data.
\vspace{-0.2cm}
\begin{findingenv}{Summary}{finding:symptom-summary}
	% Our data set reveals that data-centric API misuses exhibit symptoms similar to those observed in deep learning API misuses. 
	Program crashes are the most prevalent symptom in data-centric libraries. However, we observe that our libraries exhibit a higher proportion of misuses that result in incorrect output. 
\end{findingenv}

%%%%%%%%%%%%%%%%%%%%%%%%%%%%%%%%%%%%%%%%%%%%%%%%%%%%%%%%%%%%%%%%%%%%%%%%%%%%%%%%%%%%%%%%%%%%%%%%%%%%%%%%%%%%%%%%%%%%

\vspace{-0.3cm}
\subsection{Documented API Directives}
We find that \doccaveatsPerc of the misuses have documentation that explicitly states the correct usage of the API. 
% Table \ref{tab:misuse-doc-stats} provides a summary of the number of misuses with and without explicit API directives, and their distribution across the five libraries. 
Among the libraries that we analyzed, scikit-learn has the highest percentage of misuses with explicit API directives. 
We also analyze the relationship between misuse symptoms and the presence of explicit API directives. 
Only \checknum{20\%} of misuses resulting in low efficiency are accompanied by documented directives. In contrast, at least half of the misuses that result in unexpected output or warnings have explicitly documented API directives. These findings indicate that developers misuse \dsml{} APIs despite explicitly documented directives.

\vspace{-0.3cm}
\section{Discussion}
\label{sec:disc}

The goal of this empirical study is to understand whether the new types of API misuses, root causes, and symptoms found for deep learning libraries are specific to the deep learning domain or may extend to other libraries.
Specifically, we investigated misuses of Python data processing, machine learning, and visualization APIs and categorized them using an existing deep learning misuse taxonomy \cite{dl-api-misuse-llm:2024}.
We find that the misuses of these libraries exhibit similar behavior with those of deep learning libraries, with the heavy reliance on data being the common characteristic leading to many of the misuses.
Furthermore, the multitude of parameters in \dsml{} APIs, which accept various data types and formats, contribute significantly to the occurrence of misuses.
Our findings have implications for language and API design, surfacing information buried in documentation, and designing misuse detectors.

\vspace{0.1cm}
\noindent\textbf{\textit{Implications for language support and API design:}}
We find that \datadepPerc{} of the misuses in our data set are data dependent. Furthermore, \checknum{56\%} of these data-dependent misuses are due to violating various parameter/argument directives. For example, seaborn's \texttt{heatmap} method expects the input data in wide-format. Internally, the API checks if the input is a pandas dataframe or a rectangular array which then can be converted to a pandas dataframe. For this type checking, seaborn uses Python function \texttt{isinstance()}. Other than that, the API has no further verification to check if the input is in the correct format. 
Python does not provide any tools that can help library designers enforce such restrictions. Newer Python versions (after 3.5) did introduce type hints with the \textit{typing} module, which can then be used for static type checking by external tools such as \textit{mypy} \cite{mypy}). 
However, type checking alone cannot infer the \textit{content} of various data structures. 
While third-party library APIs typically include checks for data format (e.g., pandas dataframe or NumPy array), ensuring the integrity of data content, such as detecting mixed data types within a column of a pandas dataframe, poses a challenge in API design.
Programming languages like Java have annotation frameworks (e.g., Java Checker Framework or IntelliJ IDEA's Java Annotations) that enforce API contracts such as parameters not being null, being within a certain range, or in a specific format like an email; however, there is limited support for inter-parameter dependencies, especially if it is dependent on data. Overall, our findings imply that programming language designers need to provide tools that can be used to enforce necessary restrictions when designing \dsml{} third-party APIs. 

%% \ag{Since we do not analyze useful error messages, I think we can comment this out.}
% It is also interesting to note that in the example above, the returned error message is triggered by an internal API when it tries to convert values in the dataframe to float. The error message is generic, because it was generated by Python when the type conversion failed. As a result, it does not say anything about the dataframe being in the incorrect format. Therefore, library designers also need to provide meaningful error messages by performing necessary checks on their own instead of letting them be handled by subsequent checks in lower abstraction levels.

\vspace{0.1cm}
\noindent\textbf{\textit{Implications for API documentation:}}
Approximately half of the misused APIs did not have documented \cavs, which could potentially have led to their misuse. For example, Matplotlib's \texttt{datestr2num(...)} expects the input string to be in the date-first format (e.g., 04-21-2021) even though the documentation does not explicitly state this. The documentation, however, notes that Matplotlib uses the \textit{dateutil} library to parse a string to date. Unless developers investigate the documentation of \textit{dateutil}, it is not possible to know the assumptions made by API designers of \texttt{datestr2num(...)}.

% On the other hand, the other half of the misuses in our data set have explicitly documented \cavs{} in the target library's documentation. However, not all these \cavs{} are easily accessible. 
% We find that some of the \cavs{} are not in the API's documentation page, but rather in a different section (e.g., tutorial). For example, one of the misuses we observed was related to creating custom training loops using TensorFlow's \texttt{add\_n(...)}, where developers need to add model loss to the loss calculated between expected and predicted outputs. The documentation of \texttt{add\_n(...)} does not contain information about using it to implement a custom training loop. However, we found this \cav{} \textit{``If you are writing your own training loop,then you need to be sure to ask the model for its regularization losses.[...]This implementation works by adding the weight penalties to the model's loss[...]''} in TensorFlow's tutorial page \cite{tf-loss} buried in text.

While missing documentation or hard-to-navigate documentation is not specific to \dsml{} libraries per se~\cite{API-doc-patterns-Maalej:2013, Rec-API-Robillard:2015, API-taxo-Monperrus:2012}, we speculate that the specialized domain knowledge of using these libraries might make it more problematic as evidenced by the high misuse rate of documented \cavs{} in our data set.
Thus, we encourage \dsml{} library authors to explicitly document all \cavs{}, especially those related to expected data formats and not to assume any domain knowledge on behalf of the users.
Researchers may also help by enhancing previously proposed techniques for identifying and surfacing critical information in documentation~\cite{Rec-API-Robillard:2015}, or even augmenting documentation to include missing information~\cite{doc-aug-Treude:2016}.
Our data set of API misuses along with any documented \cavs{} can provide a starting point for further pursuing this research direction.

%\sn{I commented this out because I'm not sure what the point is.. it basically boils down to provide better documentation}
%When a misuse results in a run-time error, developers may use the error message to identify the root cause of the problem. For example, when scikit-learn raises the error message \texttt{``TypeError: Encoders require their input to be uniformly strings or numbers. Got ['float', 'str']''}, a developer could understand that the problem lies in the data format. On the other hand, \texttt{``ValueError: cannot convert string to float [...]} raised by seaborn's \texttt{heatmap(...)} API does not indicate that the input is in the incorrect format. While some of the error messages could be useful to identify the misused API, some of them are not informative about the actual problem. In an ideal world, we would expect the API to throw an error message pointing to the root cause. When that is not an option, library users have to resort to documentation (official or crowd-sourced) which has examples with explanations. seaborn's \texttt{heatmap()} documentation has an example demonstrating how to convert pandas dataframe to correct format. Therefore, future research can focus on pushing information in documentation towards developers so that potential misuses can be avoided.

\vspace{0.1cm}
\noindent\textbf{\textit{Implications for misuse detectors:}}
% Existing API misuse detectors \cite{MuDetect-Amann:2019, MisuseHint-Liang:2022, API-KG-Ren:2020, TADAF:2022} depend on analyzing the program data and control flow to detect misuses. However, data and control flow analysis will not detect problems with vector dimensions, string formats, or content in tabular data structures (e.g., pandas dataframe). 
Wei \etal{} \cite{dl-api-misuse-llm:2024} showed that existing misuse detectors are not effective in detecting deep learning API misuses. Our study shows that the characteristics of deep learning API misuses extend to data-centric APIs, and that parameter misuse and furthermore data-dependent misuses are even more prevalent in these libraries. 
To detect \dsml{} API misuses, tool developers need to not only consider dynamic analysis but to develop techniques that relate the format/content of data to the API calls and parameter values in the code.
%Developers misuse APIs even when they have explicitly documented API caveats. For example, Matplotlib provides an API \texttt{pyplot.text(...)} \cite{mpl_text} to add a text label to the plot. This API takes arguments \texttt{x, y} to position a text label in the plot. While documentation indicates \texttt{x, y} are float, it also has a textual description \textit{``The position to place the text. By default, this is in data coordinates [...]''}. As shown in Figure \ref{lst:bug_type}, the misuse instance in our data set, a developer passed a string argument to this API even though the type of \texttt{x} axis was date. Even though the caveats are right in the parameter description, not focusing on the data at hand resulted in this misuse. 
%Data dependent API misuse like this is difficult to detect unless developers are extremely attentive to the data. 
We note that tools such as \textit{Data Linter}~\cite{data-linter:2017} provides summary statistics of data and trigger notifications for incorrectly inferred types in data (e.g., string instead of numeric). While Data Linter allows developers to preprocess a data set before they feed to a program, any modifications or data manipulation that happens during the program execution is out of its scope.
%What we envision is a tool that trigger warnings when a developer accidentally passes arguments with incompatible types or format which need dynamic analysis. 
Our detailed analysis of the misuses we found and the data set we provide can help tool developers build more advanced misuse detectors.

\vspace{-0.7cm}
\section{Threats to Validity}

\textit{\textbf{Internal validity:}} Internal validity threats relate to the degree to which the study's outcomes are attributed to the methodology, minimizing personal bias.
We followed a manual labeling approach, and any manual analysis is prone to subjectivity.
To reduce subjectivity, we iteratively developed a stable coding guideline by having three annotators annotate each thread in $\sim$43\% of our \so{} thread sample. Once we stabilized the coding guideline, we continued with one annotator per data point for the GitHub commits. To further reduce subjectivity, we created minimal reproducible examples for each misuse and conservatively eliminated any misuses that we could not reproduce.

% To reduce subjectivity, we iteratively developed a clear coding guideline by having three annotators annotate each thread in $\sim$43\% of our \so{} thread sample.
% Once the coding guideline was stable, we continued with two annotators per data point for the remaining \so{}threads to also reduce subjectivity.
% \sn{what about github where it was only one reviewer? can we reduce this part a bit and get to the point that we developed a stable coding guideline through an iterative process before labeling all data?}

% When determining API misuse types, two authors used open card sorting to group misuses with similar characteristics and then assign labels to each group. To provide further validation of the group labels and assigned data to each group, the misuses and their labels were verified by a third author who was not involved with the card sorting step. 

\noindent\textit{\textbf{Construct validity:}} Construct validity threats relate to the extent to which our study reflects the concept of API misuse within the defined scope.
Approximately 8\% of the both \so{} threads and the GitHub commits that we analyzed contained API misuses.
While this ratio matches previous misuse studies~\cite{Mubench-Amann:2016,dynamic-misuse-He:2023}, we note that Wei \etal~\cite{dl-api-misuse-llm:2024} found misuses in a larger portion of their analyzed commits (21\%).
One notable differences is that the libraries that we study do not rely on hardware devices, a characteristic common in deep learning libraries. 
Additionally, we do not consider deprecation handling as API misuse.
These two categories account for approximately 58\% of the misuses found in Wei et al.'s data \cite{dl-api-misuse-llm:2024}.
We note that we did find \checknum{48} commits that fix deprecated API usage that we did not account as misuses, which would have largely increased our data set size.
Overall, our smaller total number of misuses is likely due to our stricter API misuse definition discussed in Section~\ref{sec:background}, but which is more aligned with our goal of observing misuses within the domain of \dsml{} libraries.

\vspace{0.1cm}
\noindent\textit{\textbf{External validity:}} External validity threats relate to the generalizability of our results.
We selected \lib popular Python libraries to study the misuse types in \dsml{} applications. 
We did not study all \dsml{} libraries in Python or those in other programming languages. Thus, our results may not generalize beyond the selected libraries. However, our selection represents more domain diversity than the typical deep learning focus~\cite{TF-bugs-Zhang:2018,DL-bug-char-Islam:2019,dl-api-misuse-llm:2024} and extends even beyond machine learning in general.
% These libraries cover diverse subdomains such as data processing, model creation, and data visualization. While we did not study all \dsml{} libraries in Python or those in other programming languages, our selection represents more domain diversity than the typical deep learning only focus~\cite{TF-bugs-Zhang:2018,DL-bug-char-Islam:2019,dl-api-misuse-llm:2024} and even beyond machine learning in general. 
We encourage researchers to further study more \dsml{} libraries.
We also prioritized finding commits for diverse APIs, which may mean that we missed observing more instances of the same or similar misuses.

\vspace{-0.3cm}
\section{Related work}
\label{rel}

\textit{\textbf{Types of API Misuse:}}
Amann \etal{} \cite{Mubench-Amann:2016} created a data set of Java API misuses known as MuBench. 
In followup work, the authors used this data to create a taxonomy for classifying Java API misuses (MuC) \cite{Detector-Eval-Amann:2019}. 
By analyzing API documentation of JDK, JFace, and Java Commons Collection, Monperrus \etal{} \cite{API-taxo-Monperrus:2012} developed a taxonomy of types of API directives. 
Schlichtig \etal{}~\cite{FUM:2022}'s API directive definition that we use in Section~\ref{sec:background} is inspired from Monperrus \etal{}'s work.
Schlichtig \etal~\cite{FUM:2022}'s framework for classifying API misuses is probably the most comprehensive API misuse classification taxonomy/framework, mainly derived from the Java literature.

% \ag{I commented this paragraph, and included one sentence to the previous paragraph.}

% Schlichtig \etal{} \cite{FUM:2022} created a framework for classifying API misuses (FUM) after surveying and comparing the types of misuses discussed in the literature.
% To the best of our knowledge, this is the most comprehensive API misuse classification taxonomy/framework to date since it considers all previously discussed misuses in the literature.
% Additionally, it provides clear definitions and scoping options, which we leverage to define our study scope in Section~\ref{sec:background}.

He \etal{} \cite{dynamic-misuse-He:2023} created a taxonomy (PUM) for Python misuses, comparing it to MuC and FUM above.
Their work provides valuable insights of the impact of the dynamic features of Python on using APIs.
The deep learning API misuse taxonomy by Wei \etal{} \cite{dl-api-misuse-llm:2024} is based on these previous API misuse taxonomies while extended to account for unique characteristics of the deep learning domain \cite{dynamic-misuse-He:2023,FUM:2022,Detector-Eval-Amann:2019,misuse-wild-Li:2021}.
In our study, we utilize the deep learning taxonomy to categorize misuses of \dsml{} API in general, with certain modifications made to the categories to enhance expressiveness.

\noindent\textit{\textbf{\cavs:}} A library's documentation contains information on how to use an API. While some of the information is obvious and thus has a little value \cite{API-doc-patterns-Maalej:2013}, some information is critical for the correct program behavior such as call order and condition checking~\cite{API-KG-Ren:2020}. Due to the large amount of textual details, developers could fail to notice important information when learning a new API \cite{API-doc-patterns-Maalej:2013}. To alleviate this problem, Robillard and Chhetri proposed a technique to automatically identify important knowledge items in documentation by analyzing linguistic patterns in sentences~\cite{Rec-API-Robillard:2015}. %The authors found that words such as \emph{must}, \emph{shall}, and  \emph
Monperrus \etal~\cite{API-taxo-Monperrus:2012}'s taxonomy of Java API directives is based on analyzing documentation. 
We do not systematically analyze documentation, but we look for documented \cavs{} specifically related to the misuses we observed.
Our results showed that the majority of the misused APIs (\doccaveatsPerc) have documented directives, emphasizing the need for additional support for highlighting and surfacing this information to developers.

\vspace{0.1cm}
\noindent\textit{\textbf{API misuse detection:}}
There is a long line of research that aims to detect API misuses. Early API misuse detectors mined frequent usage patterns from source code to identify correct usages; deviations from those mined patterns are then considered violations \cite{Misuse-Monperrus:2013, PR-Miner-Li:2005, RGJ-Ramanathan:2007, Colibri-ML-Lindig:2015, GrouMiner-Nguyen:2009, Jaddet-Wasylkowski:2007, MuDetect-Amann:2019}. Pattern-based techniques suffer from low precision and low recall, because these tools are limited to detecting only the frequently occurring usages. 
%Thus infrequent correct usages are mistakenly identified as misuses. 
To remedy this, researchers developed detectors to include correct usages automatically captured from documentation and client projects \cite{API-KG-Ren:2020, API-graph-Zeng:2021, MisuseHint-Liang:2022}. Additionally, some detectors use domain specific language (DSL) to detect API misuses where usage constraints are manually specified by an expert or automatically extracted from documentation \cite{Cognicrypt-Kruger:2017, IM-Gu:2019, SSL-Li:2019, gulami-thesis:2022}. 
There is also recent work that demonstrates the possibility of using large language models (LLMs) for misuse detection~\cite{dl-api-misuse-llm:2024}.
In this paper, our goal is not to design a misuse detector but to understand the nature of API misuses of \dsml{} libraries, paving the way for future research to use our findings to design and develop detection tools. For example, we find that \datadepPerc of misuses in our data set are data dependent, suggesting that tool builders need to account for the data driven nature of certain libraries when designing API misuse detectors for these libraries.

\vspace{0.1cm}
\noindent\textit{\textbf{Reasons for misusing APIs:}}
There is also a line of work that aims to understand the difficulties developers face when using APIs~\cite{API-hard-learn-Robillard:2009, API-docs-Meng:2018,Obstacles-API-learn-Hou:2011,NadiCryptoAPIs16} as well as general API usability problems~\cite{usability-Myers:2016}.
 While we do not investigate \textit{why} developers misuse APIs of \dsml{} libraries, our results (e.g., undocumented \cavs{}) suggest potential API usability issues.

\vspace{-0.3cm}
\section{Conclusion}

The idea for this empirical study started with our conjecture that many of the recently observed API misuses of deep learning libraries are not necessarily specific to machine learning or deep learning per se but are actually due to the \dsml{} nature of these libraries.
accordingly, we study API misuse in five additional \dsml{} libraries and indeed find similar API misuse characteristics.
We find that more than half of the misuses (\datadepPerc) are data-dependent and that API parameter misuse is even more prevalent than deep learning libraries.
While run-time errors are the most observed symptom, incorrect output comes as a close second.
Data-conversion errors emerge as the primary root cause of the misuses we observed.
Surprisingly, even when APIs were accompanied by explicit \cavs, developers still encountered difficulties, suggesting challenges in extracting relevant information from textual documentation.
Reflecting on our findings, we discuss implications for language and API design, misuse detectors, and surfacing buried information in documentation.
Our in-depth analysis of \dsml{} API misuses, along with the data we share containing the misuses and their (documented) \cavs, paves the way for building better support tools and misuse detectors.

% \section{Data Availability}
% All our data and analyses are available on our online artifact page \cite{dataset}.
\pagebreak
\bibliographystyle{ACM-Reference-Format}
\bibliography{references}

\end{document}